\documentclass[sigconf, nonacm]{acmart}

\usepackage[table, dvipsnames]{xcolor}

\usepackage{amsthm}
\usepackage{xspace, xfrac}
\usepackage{enumitem}
\usepackage[caption=false]{subfig}
\usepackage{graphicx}
\usepackage{booktabs}
\usepackage{comment}

\usepackage{tabularx, multirow}
\newcolumntype{L}{>{\raggedright\arraybackslash}X}
\newcolumntype{C}{>{\centering\arraybackslash}X}
\newcolumntype{R}{>{\raggedleft\arraybackslash}X}

\newcommand{\NAME}{VIPIR\xspace}
\newcommand{\mathpolyring}{\mathcal{R}_q}
\newcommand{\polyring}{$\mathpolyring$\xspace}
\newcommand{\mathDB}{\mathtt{DB}}
\newcommand{\DB}{$\mathDB$\xspace}

\newcommand{\circled}[1]{\textbf{\large \textcircled{\small #1}}}

\begin{document}

\title{VIPIR: A Versatile GPU Framework for Integrating Private Information Retrieval Protocols}

\author{Jongmin Kim}
\orcid{0000-0003-2937-3073}
\authornote{Both authors contributed equally to this research.}
\authornote{Part of this work was conducted during an internship at NVIDIA.}
\affiliation{%
  \institution{Seoul National University}
  \city{Seoul}
  \country{South Korea}
}
\email{jongmin.kim@snu.ac.kr}

\author{Hyesung Ji}
\orcid{0009-0009-9288-159X}
\authornotemark[1]
\affiliation{%
  \institution{Seoul National University}
  \city{Seoul}
  \country{South Korea}
}
\email{kevin5188@snu.ac.kr}

\author{Jean-Luc Watson}
\orcid{0000-0001-9175-9415}
\affiliation{%
  \institution{NVIDIA}
  \city{Santa Clara}
  \state{CA}
  \country{USA}
}
\email{jwatson@nvidia.com}

\author{Charles Gouert}
\orcid{0000-0002-7670-830X}
\affiliation{%
  \institution{NVIDIA}
  \city{Santa Clara}
  \state{CA}
  \country{USA}
}
\email{cgouert@nvidia.com}

\author{G. Edward Suh}
\orcid{0000-0001-6409-9888}
\affiliation{%
  \institution{NVIDIA / Cornell University}
  \city{Westford}
  \state{MA}
  \country{USA}
}
\email{esuh@nvidia.com}

\author{{Jung Ho} Ahn}
\orcid{0000-0003-1733-1394}
\affiliation{%
  \institution{Seoul National University}
  \city{Seoul}
  \country{South Korea}
}
\email{gajh@snu.ac.kr}


\begin{abstract}

While private information retrieval (PIR) enables private database services by fully concealing access patterns, it simultaneously requires high computational throughput, large memory capacity, and substantial memory bandwidth.
We introduce \NAME, a versatile GPU framework that co-designs PIR protocols with GPU acceleration.
We develop a unified analytic model showing that state-of-the-art PIR protocols fall into two categories with complementary limitations, and propose two protocols that flexibly combine techniques across these categories, overcoming the limitations of both classes.
These protocols incorporate a GPU-friendly data compression method called expansion-based ring packing (ExpPack), which offers a high degree of parallelism and minimal communication cost.
\NAME applies further optimizations to core operations, including number-theoretic transforms (NTTs) and various matrix-matrix multiplications (GEMMs).
Notably, we develop a tensor-core-based execution method for database multiplication by interpreting it as a mixed-integer-type GEMM.
We also design memory-efficient scheduling methods that minimize intermediate buffers and enable multi-GPU scaling under memory capacity constraints.
Overall, \NAME achieves orders-of-magnitude higher throughput than prior PIR systems while reducing communication and memory overheads, making large-scale PIR practical.

\end{abstract}

\maketitle 

\section{Introduction}
\label{sec:intro}

As data privacy emerges as a key challenge for digital services, private information retrieval (PIR)~\cite{jacm-1998-chorpir, sfcs-1997-computational-pir} has become a critical cryptographic primitive.
PIR enables a client to retrieve a record from a database (\DB) hosted on a remote server without revealing which record is being accessed.
This capability is increasingly relevant in practice: for example, the Ethereum Foundation is exploring PIR to fully anonymize blockchain access~\cite{ethereum-pir}, while Apple employs PIR for privacy-preserving contact discovery and safe browsing~\cite{AppleHE}.

Despite its theoretical appeal, deploying PIR at scale remains challenging due to a fundamental requirement: to preserve privacy, the server must perform a linear scan over the entire \DB per query.
Single-server PIR~\cite{sfcs-1997-computational-pir} protocols execute homomorphic encryption (HE)~\cite{2021-standard} operations over \DB during this scan; HE allows direct computations on encrypted data, enabling private record retrieval without revealing the accessed index.
This HE-based linear scan imposes stringent system demands, requiring high computational throughput, large memory capacity, and substantial memory bandwidth simultaneously.

GPUs offer the massive parallelism needed to accelerate such scans.
Compared to server-class CPU systems, which have been the primary platform for most prior cryptographic PIR studies, modern GPUs provide significantly higher memory bandwidth via High Bandwidth Memory (HBM) DRAM and orders-of-magnitude greater computational throughput at lower per-operation energy cost.

\begin{figure}
    \centering
    \includegraphics[width=0.99\linewidth]{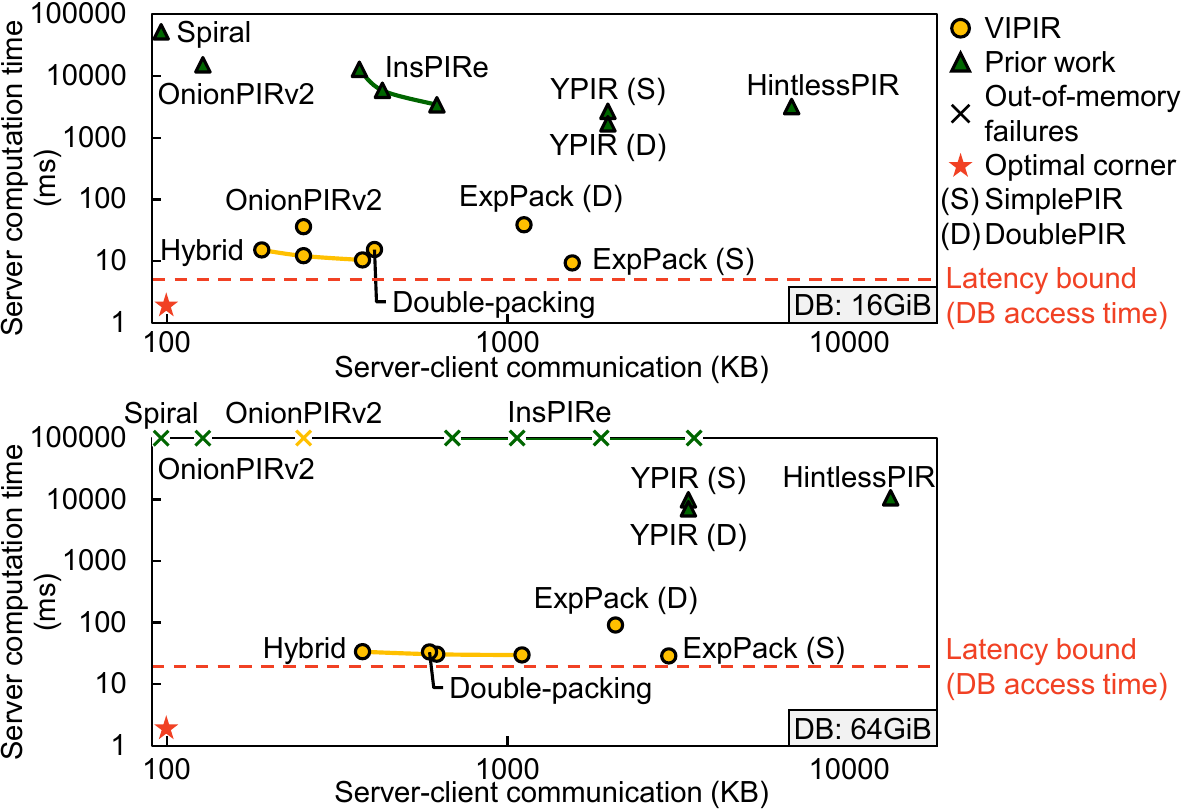}
    \Description{}
    \caption{Server-side online computation time and client-server communication of prior CPU-based PIR studies and \NAME for 16GiB/64GiB \DB (single batch). \NAME supports multiple protocols (hybrid, double-packing, OnionPIRv2, ExpPack (S/D)).
    InsPIRe and our hybrid protocol span multiple \DB dimensionalities.
    Refer to Table~\ref{tab:eval:vs_cpu} for more details. 
    }
    \label{fig:intro}
\end{figure}

Yet, only a few GPU-focused studies on PIR exist~\cite{ccs-2025-shiftpir, github-PIRonGPU, hpca-2026-ive}, and they focus on a single protocol with little regard for adapting protocol design to efficient GPU execution.
Although GPUs have relatively limited DRAM capacity compared to CPU systems, which can provision TBs of memory, PIRonGPU~\cite{github-PIRonGPU} and the GPU implementation of IVE~\cite{hpca-2026-ive} adopt protocols that expands the \DB size by $\sim$4$\times$.
This \emph{\DB bloating} becomes a critical bottleneck for scalability.

For a principled approach, we analyze existing PIR protocols and develop a unified analytical model that enables reasoning about them from a common perspective.
Despite diverse proposals, we observe that most PIR protocols fall into two categories under this model: one (which we term scalar-HE approach) with high client-server communication overhead and the other (poly-HE approach) with \DB bloating.



In this paper, we present \textbf{\NAME}, a versatile GPU framework enabling GPU-aware PIR protocol design and optimized execution on NVIDIA GPUs.
Beyond accelerating existing PIR protocols on GPUs, we derive new protocols from our unified model that are inherently more amenable to efficient GPU execution.
These protocols center on an HE primitive called \emph{ring packing} (or simply packing)~\cite{asia-2017-column-ring-pack, crypto-2023-hermes, sp-2021-pegasus, acns-2021-cdks} that compresses encrypted data to substantially smaller sizes (0.16\% in our settings), trading communication for added computations.
While packing has been used in prior PIR work~\cite{usenixsec-2024-ypir, crypto-2024-hintless, sosp-2023-tiptoe, iacr-2025-inspire}, we discover its different usage in our protocols and also develop a novel expansion-based packing (ExpPack) method.

In particular, we introduce (i) a \emph{double-packing} design applying ring packing twice to significantly improve compression, and (ii) a \emph{hybrid} design repurposing ring packing to transition between scalar-HE and poly-HE approaches.
This avoids the \DB bloating problem of poly-HE, while alleviating the communication overhead of scalar-HE.


ExpPack augments these protocols by replacing the transfer of $n=1{,}280$ ``special'' ciphertexts (120MiB) with a single ciphertext (96KiB) that is expanded on the server side, leveraging GPUs’ high computational capability to trade communication for additional computation.
While prior studies also propose communication-efficient packing methods, they rely on sequential procedures or memory-intensive precomputation that limit GPU efficiency~\cite{iacr-2025-inspire, crypto-2024-hintless}.
In contrast, ExpPack is designed to expose massive data parallelism through thousands of independent sub-operations---namely number-theoretic transforms (NTTs) and modular matrix-matrix multiplications (GEMMs)---enabling effective utilization of hundreds of streaming multiprocessors.

\NAME incorporates various architecture-aware GPU optimizations to achieve practical performance.
Notably, we leverage NVIDIA tensor cores~\cite{ieee-micro-2018-tensorcore} to accelerate core INT8-INT32 GEMMs by emulating them on the INT8 GEMM datapath, enabled by co-designing parameters and protocols with the GPU implementation.
We also adjust data layouts for contiguous memory access, apply numerical techniques such as lazy reduction~\cite{tches-2021-100x, asplos-2026-cheddar} to reduce instruction count, and minimize temporary data by carefully scheduling jobs under constrained GPU memory.
\NAME also supports multi-GPU execution by combining \DB sharding with batch-wise job distribution, enabling low-latency scaling to larger databases.

\NAME delivers substantial performance gains while supporting \DB sizes up to 64GiB on a single GPU (H100 NVL 94GB), while avoiding the out-of-memory failures common in prior work due to excessive memory usage.
As shown in Figure~\ref{fig:intro}, \NAME achieves orders-of-magnitude lower latency than prior work, with our co-designed double-packing and hybrid protocols simultaneously reducing communication costs.
For a 64GiB \DB, these protocols attain execution times of 33.0--33.5ms, only 1.67--1.69$\times$ above the 19.7ms hard latency bound for streaming \DB once from DRAM, indicating that our implementation closely approaches the theoretical performance limit.
\NAME also achieves high throughput of 163.1--232.8 queries per second (QPS), outperforming prior GPU-based work, ShiftPIR~\cite{ccs-2025-shiftpir}, by 652.6--931.3$\times$.

We summarize the main contributions of \NAME as follows:
\begin{itemize}[leftmargin=*]
\item We develop a unified analytical model that captures diverse PIR protocols, exposes their limitations, and enables new double-packing and hybrid designs.
\item We introduce ExpPack, a GPU-friendly expansion-based ring packing method that exposes massive parallelism while minimizing client-server communication.
\item We build optimized GPU implementations of PIR with tensor-core acceleration, memory-aware layout and scheduling, numerical optimizations, and multi-GPU scaling.
\end{itemize}
\section{Motivation \& Background}
\label{sec:back}

Vectors (e.g., $\mathbf{v}$) and matrices (e.g., $\mathbf{A}$) are denoted in boldface; all vectors are column vectors.
An $M\times K$ by $K\times N$ matrix-matrix multiplication is denoted as $M\times N\times K$ GEMM.
Major notations and symbols are summarized in Appendix~\ref{app:symbol}.

\subsection{Challenges of private information retrieval (PIR)}
\label{sec:back:pir}

Dozens of PIR protocols~\cite{ccs-2024-kspir, sp-2022-spiral, sp-2018-sealpir, ccs-2024-respire, popets-2016-xpir, ccs-2021-onionpir, iacr-2024-whispir, iacr-2025-inspire, usenixsec-2024-ypir, popets-2023-frodopir, iacr-2025-onionpirv2, sosp-2023-tiptoe, crypto-2024-hintless} exhibit intricate trade-offs across client/server-side computation, memory access, storage overhead, and client-server communication.
Nevertheless, most constructions build upon prior work, reusing core components and design principles from earlier protocols.

Without extensive client-side preprocessing~\cite{sp-2024-piano}, PIR protocols require the server to linearly scan the entire database (\DB) per query to hide the accessed record.
The resulting uniform access requirements to GBs of data challenge computing systems to deliver high bandwidth and large capacity simultaneously, while the memory hierarchy offers limited assistance in alleviating this bottleneck.

In this paper, we distill common factors of PIR to build a principled PIR framework, rethinking how modern accelerators such as GPUs reshape the design space.
To focus our analysis, we exclude PIR protocols relying on restrictive assumptions, such as multi-server non-collusion~\cite{jacm-1998-chorpir}, single-client batch querying~\cite{sp-2023-vectorized, sp-2018-sealpir}, or extensive client-side \DB preprocessing~\cite{sp-2024-piano}, which hinder practical adoption.


\begin{figure*}
    \centering
    \includegraphics[width=0.99\linewidth]{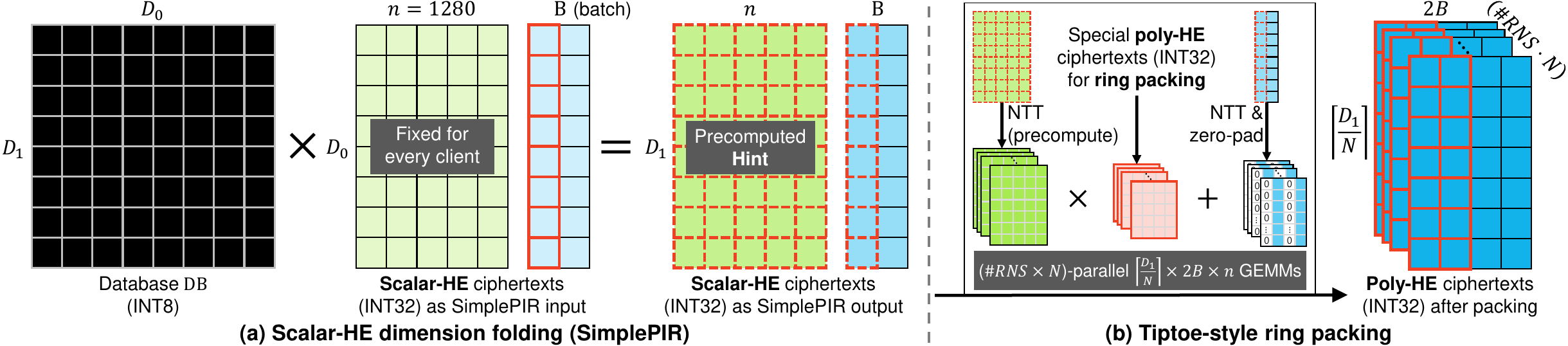}
    \Description{}
    \caption{(a) Server-side $B$-batch execution of SimplePIR~\cite{usenixsec-2023-simplepir} and (b) ring packing based on Tiptoe~\cite{sosp-2023-tiptoe}. Each cell represents an INT8/INT32 element. Red-bordered cells mark client-exchanged data, where dotted cells are not exchanged under packing.}
    \label{fig:simplepir}
\end{figure*}

\subsection{Homomorphic encryption (HE)}
\label{sec:back:he}

HE allows direct computation on encrypted data, which is vital for implementing PIR.
For HE ciphertexts $\mathtt{ct}_1$ and $\mathtt{ct}_2$ respectively encrypting plaintexts $\mathtt{pt}_1$ and $\mathtt{pt}_2$, HE schemes have the following linear property, allowing inter-ciphertext additions and plaintext-ciphertext multiplications:
\begin{equation}
\label{eq:he-linear}
\mathtt{pt}_0 \cdot \mathtt{ct}_1 + \mathtt{ct}_2 \xrightarrow[]{\text{decrypt}} \mathtt{pt}_0 \cdot \mathtt{pt}_1 + \mathtt{pt}_2
\end{equation}

We classify various HE schemes into two categories:
\begin{itemize}[leftmargin=*]
    \item \textbf{Scalar-HE} encrypts an integer (plaintext) into a length-$(n+1)$ integer vector (ciphertext).
    \item \textbf{Poly-HE} encrypts an $N$-coefficient integer polynomial (plaintext) into two or more polynomials in the ring $\mathcal{R}_q=\mathbb{Z}_q[X]/(X^N+1)$ (ciphertext).
\end{itemize}

Specifically, scalar-HE refers to the basic Regev encryption based on the learning-with-errors (LWE) problem~\cite{jacm-2009-regev}.
We use parameter settings, where we can encrypt an 8-bit integer (INT8) into a 32-bit integer (INT32) vector.
Poly-HE includes various HE schemes based on ring LWE (RLWE)~\cite{eurocrypt-2010-rlwe}, ring GSW (RGSW)~\cite{crypto-2013-gsw}, or module LWE~\cite{dcc-2015-mlwe}.
%
%
%
For RLWE, which is most commonly used, each ciphertext comprises two polynomials in $\mathcal{R}_q$; i.e., $\mathtt{ct} \in \mathcal{R}_q^2$.
We use $n=1{,}280$, $N=2^{12}$, $q\simeq2^{87}$ throughout this work (detailed in \S\ref{sec:impl}).

\subsection{Scalar-HE arithmetic vs. poly-HE arithmetic}
\label{sec:back:he-types}

Computations with scalar-HE are relatively simple without the need for modular arithmetic~\cite{usenixsec-2023-simplepir}.
For our typical parameters, plaintext-ciphertext multiplications and inter-ciphertext additions respectively translate to regular INT8-INT32 scalar-vector multiplications and INT32 vector additions.

Handling polynomials for poly-HE is computationally more demanding.
A polynomial in \polyring has $N$ coefficients in $\mathbb{Z}_q$, where
residue number system (RNS) is used as an optimization to decompose the modulus $q$ into $\text{\#RNS}$ number of small primes.
For example, we set $q=q_0\cdot q_1 \cdot q_2$ ($\text{\#RNS}=3$) with $q_i<2^{32}$, converting a polynomial into a $3\times N$ INT32 matrix, where each row corresponds to a different $q_i$:
%
\begin{equation*}
    a_0+a_1X+\cdots+a_{N-1}X^{N-1} \rightarrow \begin{pmatrix}a_0\ \%\ q_0 & \cdots & a_{N-1}\ \%\ q_0\\ a_0\ \%\ q_1 & \cdots & a_{N-1}\ \%\ q_1\\ a_0\ \%\ q_2 & \cdots & a_{N-1}\ \%\ q_2\\\end{pmatrix}
\end{equation*}
While poly-HE inter-ciphertext additions become simple point-wise modulo-$q_i$ matrix additions, plaintext-ciphertext multiplications involve polynomial multiplications.

The number-theoretic transform (NTT) is a standard technique for accelerating polynomial multiplications.
For the $3 \times N$ matrix, NTT applies a variant of the Fourier transform (computational complexity is $\mathcal{O}(N \log N)$) to each row.
Once transformed, a polynomial multiplication ($\mathcal{O}(N^2)$) reduces to an element-wise multiplication modulo $q_i$ ($\mathcal{O}(N)$).


\subsection{Ciphertext bloating \& plaintext bloating}

Each HE category faces a distinct data-size blowup: \emph{ciphertext bloating} for scalar-HE and \emph{plaintext bloating} for poly-HE.
Scalar-HE encryption incurs huge data expansion of $(n+1)\cdot\frac{\text{INT32}}{\text{INT8}} = 5{,}124\times$, which we refer to as ciphertext bloating.
In contrast, poly-HE
can exploit the $N$ coefficients in a plaintext polynomial ($\in\mathcal{R}_p$, $p=2^{18}$ in this work) to encrypt $N\log_2 p$ bits of data per ciphertext.
In particular, RLWE ciphertexts ($\in\mathcal{R}_q^2$) incurs only $2\cdot\frac{\log_2 q}{\log_2 p}\times$ expansion, which is under 10$\times$.

Instead, in poly-HE, plaintexts grow in size as we apply NTT to enable efficient plaintext-ciphertext multiplications.
During this process, plaintexts in $\mathcal{R}_p$ are lifted to $\mathcal{R}_q$, resulting in a size expansion of $\frac{\log_2 q}{\log_2 p}\times$, typically around $4\times$.
This plaintext bloating issue does not arise in scalar-HE, where plaintexts (scalars) are kept in their original form.



\subsection{Exemplar PIR protocol: SimplePIR}
\label{sec:back:simplepir}

As a concrete example, we look at SimplePIR~\cite{usenixsec-2023-simplepir}, which organizes \DB as a $D_1 \times D_0$ matrix.
To query index $i^*$ along the $D_0$ dimension, the client sends $D_0$ scalar-HE ciphertexts, where only the $i^*$-th ciphertext encrypts one and the others encrypt zero, forming an encrypted one-hot representation of $i^*$.
The ciphertexts together form a $D_0 \times (n+1)$ matrix, with which the server multiplies \DB, producing a $D_1 \times (n+1)$ matrix that is interpreted as $D_1$ scalar-HE ciphertexts encrypting the $i^*$-th column of \DB.

SimplePIR exploits the fact that $D_0 \times n$ input elements are query-independent and can be reused across clients.
Consequently, most of the \DB GEMM can be precomputed, yielding a $D_1 \times n$ \textbf{hint}.
If the client system receives and stores the hint within an \emph{offline} communication phase, the client can only send $D_0$ elements per query and receive $D_1$ elements in response during the \emph{online} phase.


Figure~\ref{fig:simplepir}(a) illustrates the server-side computations for SimplePIR with multi-client batching~\cite{hpca-2026-ive}.
With hint precomputation, the server's online computation only includes a $D_1 \times B \times D_0$ INT8-INT32 GEMM for a batch of $B$ queries.

\subsection{Ring packing for hintless SimplePIR}
\label{sec:back:rp}

Numerous studies~\cite{usenixsec-2024-ypir, crypto-2024-hintless, sosp-2023-tiptoe, iacr-2025-inspire} have explored \emph{hintless} variants of SimplePIR by applying \emph{ring packing} (or simply \emph{packing}),
%
which converts $N$ scalar-HE ciphertexts with $N \times (n+1)$ elements into a single poly-HE (RLWE) ciphertext with $2 \times (\text{\#RNS} \cdot N)$ elements.
This represents a \emph{compression factor} (lower is better) of $\frac{2\cdot\text{\#RNS}}{n+1}\times$, mitigating ciphertext bloating in scalar-HE.
\#RNS can be further reduced to one after packing (referred to as modulus reduction~\cite{focs-2011-modulus-reduction}), achieving $\frac{2}{n+1}\times$  (\textless0.16\% for our parameters) compression.
This eliminates the need for hint reuse, instead enabling online communication via compact poly-HE ciphertexts.

Figure~\ref{fig:simplepir}(b) illustrates an exemplar packing based on Tiptoe~\cite{sosp-2023-tiptoe}, which involves NTTs and parallel GEMMs with additional ``special'' poly-HE ciphertexts.
Ring packing will be analyzed in detail in \S\ref{sec:unify:ExpPack}.

\subsection{GPU execution model}
\label{sec:back:gpu}

Modern NVIDIA GPUs are massively parallel systems composed of multiple streaming multiprocessors (SMs), each integrating execution units such as INT32 cores for integer arithmetic and tensor cores for high-throughput low-bit (e.g., INT8) GEMMs.
GPU kernels follow the single-instruction, multiple-thread (SIMT) model, where threads are grouped into warps (groups of 32 threads) and scheduled to SMs.
Performance is governed by occupancy, which is the number of active warps per SM relative to the hardware limit, and efficient utilization of the memory hierarchy, spanning per-SM memory resources (registers, shared memory, and L1 cache) and global memory resources (L2 cache and DRAM).

Recent GPUs employ High Bandwidth Memory (HBM) to deliver massive DRAM bandwidth (e.g., 3.94TB/s in H100 NVL 94GB) and provide high-bandwidth interconnects such as NVLink for multi-GPU configurations, enabling efficient parallelization with modest communication overhead.





\section{A Unified Approach to PIR Protocols}
\label{sec:unify}
\subsection{PIR as a compression process}
\label{sec:unify:compression}

We present a distinct perspective on PIR by interpreting it as a compression process.
A trivial PIR protocol has a server return an entire \DB as a response to each query, whereas practical protocols compress the response via HE operations to retain only the requested records.
For example, SimplePIR (\S\ref{sec:back:simplepir}) achieves a compression factor of $\frac{4(n+1)}{D_0}\times$, compressing \DB ($D_1\times D_0$ INT8 values) into a compact response ($D_1\times(n+1)$ INT32 values including the hint).


More generally, PIR protocols apply a sequence of \emph{dimension folding}~\cite{ccs-2024-kspir, sp-2022-spiral} (or simply \emph{folding}) steps to progressively compress \DB.
For a \DB with $D_k\times D_{k-1} \times \cdots \times D_0$ records, each folding step removes a $D_i$ factor from the response.


\subsection{A unified analytic model for PIR}
\label{sec:unify:model}

We observe a dichotomy in PIR protocols, where most PIR protocols rely on a single HE category for dimension folding.
More details on individual protocols are provided in \S\ref{sec:related}.

\noindent $\bullet$\ \textbf{Scalar-HE folding.}
Building upon SimplePIR, these protocols have one or two folding steps using one-hot representations (\S\ref{sec:back:simplepir}); the two-folding variant is referred to as DoublePIR~\cite{usenixsec-2023-simplepir}.
The second folding requires \emph{reinterpretation} beforehand, which treats existing data as a stream of scalars (plaintexts),
because scalar-HE does not support inter-ciphertext multiplications.
Hintless variants add a ring packing step at the end (\S\ref{sec:back:rp}).
This PIR category includes SimplePIR/DoublePIR~\cite{usenixsec-2023-simplepir}, Tiptoe~\cite{sosp-2023-tiptoe}, HintlessPIR~\cite{crypto-2024-hintless}, and YPIR~\cite{usenixsec-2024-ypir}. 

\noindent $\bullet$\ \textbf{Poly-HE folding.} An optimized protocol of this category, OnionPIRv2~\cite{iacr-2025-onionpirv2}, organizes \DB as a $2 \times 2 \times \cdots \times 2 \times D_0$ structure, utilizing RLWE plaintext-ciphertext multiplications for the initial $D_0$-dimension folding and RLWE-RGSW external products for the rest~\cite{ccs-2021-onionpir}.
Although alternative methods, such as RLWE inter-ciphertext multiplications~\cite{sp-2018-sealpir}, rotation-based encrypted matrix-vector multiplications~\cite{ccs-2024-kspir}, or MLWE operations~\cite{sp-2022-spiral}, are also used, they offer similar computation-communication trade-offs.
This category includes XPIR~\cite{popets-2016-xpir}, MulPIR~\cite{security-2021-mulpir}, WhisPIR~\cite{iacr-2024-whispir}, OnionPIR~\cite{ccs-2021-onionpir}, OnionPIRv2~\cite{iacr-2025-onionpirv2}, Spiral~\cite{sp-2022-spiral}, Respire~\cite{ccs-2024-respire}, and KsPIR~\cite{ccs-2024-kspir}.


\begin{figure}
    \centering
    \includegraphics[width=0.99\linewidth]{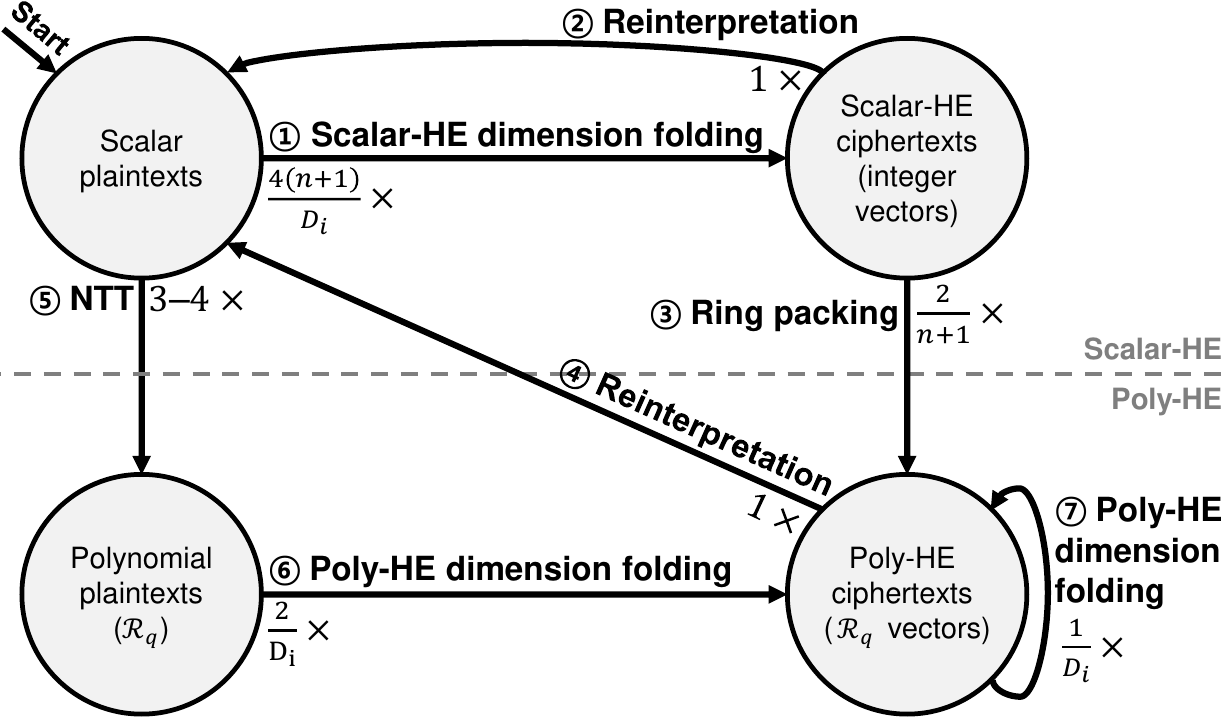}
    \Description{}
    \caption{Our unified PIR model. Each arrow, annotated with its typical compression factor (lower is better), denotes an operation that directly transforms the response state; the exact compression factor may vary depending on parameters and implementation. Here, $D_i$ denote the \DB dimensions and $n=1{,}280$ is the LWE degree.}
    \label{fig:unify}
\end{figure}

We develop a unified analytic model that overcomes the dichotomy and provides a comprehensive viewpoint for PIR protocols (see Figure~\ref{fig:unify}).
In this model, the two PIR categories follow these paths (optional paths enclosed in parentheses):
\begin{itemize}[leftmargin=*]
    \item \textbf{Scalar-HE folding:} \circled{1} $\rightarrow$ (\circled{2} $\rightarrow$ \circled{1}) $\rightarrow$ (\circled{3}).
    \item \textbf{Poly-HE folding:} \circled{5} $\rightarrow$ \circled{6} $\rightarrow$ \circled{7} $\rightarrow$ \circled{7} $\rightarrow$  $\cdots$ $\rightarrow$ \circled{7}.
\end{itemize}

There are some exceptions; for instance, SealPIR~\cite{sp-2018-sealpir} follows \circled{5} $\rightarrow$ \circled{6} $\rightarrow$ \circled{4} $\rightarrow$ \circled{5} $\rightarrow$ \circled{6}, although such designs are generally less efficient compared to other poly-HE folding approaches~\cite{ccs-2021-onionpir}.
Also, InsPIRe~\cite{iacr-2025-inspire} develops protocols that follow similar paths to our new protocols in \S\ref{sec:unify:new-protocol}, but InsPIRe exhibits significant limitations when targeting GPUs (\S\ref{sec:unify:vs-others}).

\subsection{Limitations of the two PIR categories}
\label{sec:unify:limit}

Under this model, we identify the limitations of each approach.
Scalar-HE folding suffers from large response sizes.
For \circled{1} $\rightarrow$ \circled{2} $\rightarrow$ \circled{1} $\rightarrow$ \circled{3}, the overall compression factor is:
\[
\frac{4(n+1)}{D_0}\times1\times\frac{4(n+1)}{D_1}\times\frac{2}{n+1}=\frac{32(n+1)}{D_0D_1}.
\]
Although constant terms vary with the parameter choice, the $(n+1)$ ($\ge2^{10}$) term, which stems from ciphertext bloating in scalar-HE, leads to substantial response expansion relative to the original record size.

In contrast, poly-HE folding achieves a more favorable compression factor ($\simeq \frac{8}{\prod_i D_i}$) but suffers from \emph{\DB bloating}, i.e., plaintext expansion due to the initial NTT (\circled{5}).
The resulting $\sim$4$\times$ increase in \DB size severely limits the applicability and scalability of PIR protocols for large databases.

The \DB bloating problem is particularly critical for GPUs, whose memory capacity (up to hundreds of GBs per device) is more limited than that of server-class CPUs (up to TBs).
As PIR requires a linear scan over the entire \DB, keeping \DB in GPU DRAM is essential to maximize bandwidth.
However, for real-world GB- to TB-scale databases, the additional $\sim$4$\times$ storage overhead becomes a major bottleneck, either incurring costly host-GPU communication or requiring additional GPUs merely for memory capacity.


\begin{figure}
    \centering
    \includegraphics[width=0.99\linewidth]{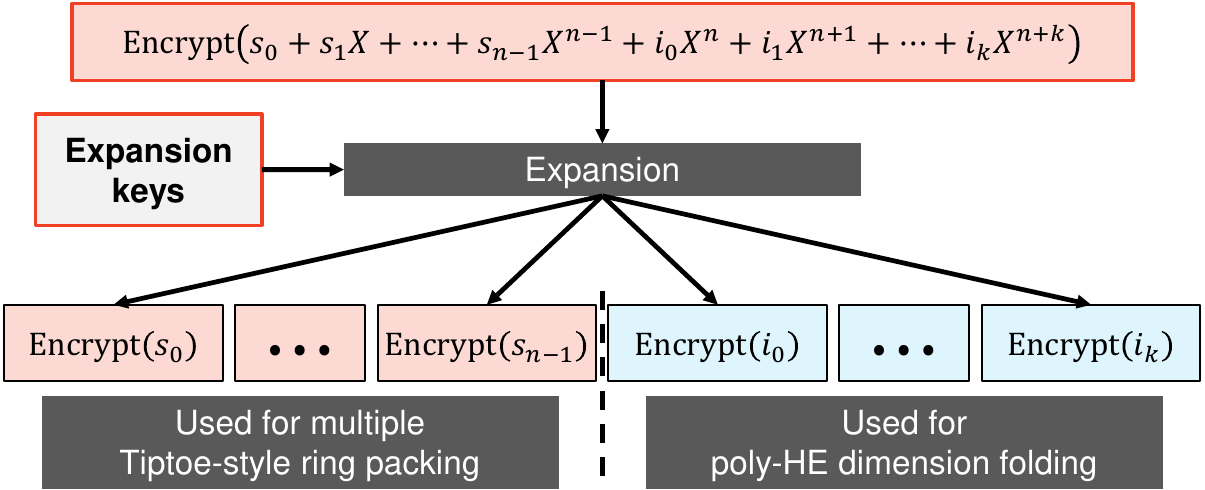}
    \Description{}
    \caption{A black-box view of expansion-based ring packing (ExpPack) used for multiple paths in the unified model; red-bordered boxes denote client-sent data.}
    \label{fig:ExpPack}
\end{figure}

\subsection{GPU-friendly protocol constructions}
\label{sec:unify:new-protocol}

From our unified model, we derive two PIR protocols that achieve improved compression for low client-server communication while avoiding \DB bloating, which is critical for GPUs, by starting with scalar-HE folding (\circled{1}):

\noindent $\bullet$\ \textbf{Double-packing:} \circled{1} $\rightarrow$ \circled{3} $\rightarrow$ \circled{4} $\rightarrow$ \circled{1} $\rightarrow$ \circled{3}.
Applying packing twice reduces the overall compression factor to $\frac{64}{D_0D_1}$, eliminating the $(n+1)$ factor.
This essentially trades off reduced communication with additional computations for performing packing twice.
We mitigate the computational overhead with a novel ring packing method (ExpPack, \S\ref{sec:unify:ExpPack}) that is especially effective for multiple packing computations.

\noindent $\bullet$\ \textbf{Hybrid:} \circled{1} $\rightarrow$ \circled{3} $\rightarrow$ \circled{7} $\rightarrow$ \circled{7} $\rightarrow$ $\cdots$ $\rightarrow$ \circled{7}.
While prior scalar-HE folding approaches used packing primarily for compression, we repurpose it as a bridge to enable hybrid PIR protocols that transition from scalar-HE folding in the initial dimension to poly-HE folding in subsequent dimensions.
For poly-HE folding (\circled{7}), we adopt RGSW-RLWE external products~\cite{jc-2020-tfhe}, as used in OnionPIR~\cite{ccs-2021-onionpir, iacr-2025-onionpirv2}.

The overall computational cost of both protocols is dominated by the first two steps (\circled{1} $\rightarrow$ \circled{3});
beyond this point, the data is sufficiently compressed ($\frac{8}{D_i}\times$) such that subsequent computations contribute minimally.



\subsection{Expansion-based ring packing (ExpPack)}
\label{sec:unify:ExpPack}

Our protocols are particularly GPU-friendly in part because of a novel expansion-based packing (ExpPack) method (\circled{3}) that drastically reduces client-server communication while exposing massive GPU parallelism.
Across existing packing methods~\cite{acns-2021-cdks, sp-2021-pegasus, asia-2017-column-ring-pack, crypto-2023-hermes}, the core idea is the same: $N$ scalar-HE ciphertexts are converted into a single poly-HE (RLWE) ciphertext using additional client-provided ``special'' ciphertexts (see Figure~\ref{fig:simplepir}(b)) and/or \emph{keys}~\cite{iacr-2025-inspire, acns-2021-cdks, crypto-2024-hintless}.

ExpPack differs primarily in how the special ciphertexts are constructed, enabling the server to expand them on-the-fly.
Expansion is an HE algorithm that enables extracting RLWE ciphertexts each encrypting $a_i$ ($0 \le i < N$) from an RLWE encryption of $a_0+a_1X+\cdots+a_{N-1}X^{N-1}$~\cite{sp-2018-sealpir, ccs-2021-onionpir}, which we use in a black-box manner in this work.

We propose coupling this expansion with the packing method from Tiptoe~\cite{sosp-2023-tiptoe}, which is otherwise impractical due to excessive communication overhead for the special ciphertexts.
For a client's length-$n$ secret $\mathbf{s}=(s_0, s_1,\cdots, s_{n-1})$, Tiptoe requires $n$ special RLWE ciphertexts for ring packing:
\[
\mathrm{Encrypt}(s_0), \mathrm{Encrypt}(s_1), \cdots, \mathrm{Encrypt}(s_{n-1}).
\]
Their total size reaches 120MiB for our parameters, orders of magnitude larger than the actual index and response ciphertexts (on the order of KBs), making Tiptoe highly impractical.

Instead, we make the client send a single RLWE ciphertext (96KiB) that encrypts all the secret values as:
\[
\mathtt{ct}_\text{ExpPack} \gets \mathrm{Encrypt}(s_0 + s_1 X + \cdots + s_{n-1} X^{n-1}), \quad n < N.
\]
Then, the special ciphertexts can be recovered from $\mathtt{ct}_\text{ExpPack}$ through expansion.
More cryptographic details of ExpPack are provided in Appendix~\ref{app:exppack_details}.

ExpPack thus drastically reduces the communication overhead for the special ciphertexts, at the cost of additional \emph{expansion keys} (up to 5.16 MiB under our parameters).
These keys, however, can be stored at the server and reused across multiple queries from the same client over time.

\subsubsection{Multi-purpose expansion}
\label{sec:unify:ExpPack:multi}

ExpPack is particularly well-suited for the protocols in \S\ref{sec:unify:new-protocol} as it can be used for multiple paths in our unified model (see Figure~\ref{fig:ExpPack}).
For the double-packing protocol, the expansion is performed only once, after which the recovered special ciphertexts can be reused for multiple packing steps.
For the hybrid protocol, we exploit the fact that $n < N$ to embed additional information in $\mathtt{ct}_\text{ExpPack}$: we embed both the $s_i$ values for packing and the index information for subsequent poly-HE folding steps.
This allows a single expansion process to serve both packing (\circled{3}) and poly-HE folding (\circled{7}) paths.

\subsubsection{Massive parallelism}
ExpPack unlocks the massive parallelism inherent in Tiptoe-style packing, which was previously prohibitive due to its high communication overhead.
After the expansion, whose acceleration via parallelization has been extensively studied in IVE~\cite{hpca-2026-ive}, thousands of independent NTTs and INT32 GEMMs follow (see Figure~\ref{fig:simplepir}(b)), exposing massive parallelism well-suited for GPUs.

Revisiting the $B$-batch SimplePIR in Figure~\ref{fig:simplepir}, the $D_1 \times n$ and $D_1 \times B$ SimplePIR outputs are first transformed into polynomials by treating each length-$N$ vector along the $D_1$ dimension as a polynomial in \polyring, with integer elements interpreted as coefficients.
This yields $\lceil D_1 / N \rceil \times n$ and $\lceil D_1 / N \rceil \times B$ polynomials, to which NTTs are applied independently.

Let $\mathbf{H}$ and $\mathbf{G}$ denote these two polynomial matrices after the NTTs.
The computation of $\mathbf{H} \cdot \mathbf{C} + (\mathbf{0} \mid \mathbf{G})$ follows, where $\mathbf{C}$ is an $n \times 2B$ polynomial matrix corresponding to expanded special ciphertexts from $B$ clients.
As RNS and NTT transform a polynomial multiplication into point-wise multiplications over $(\text{\#RNS} \cdot N)$ coefficients, this reduces to $(\text{\#RNS} \cdot N)$-parallel $\lceil D_1 / N \rceil \times 2B \times n$ INT32 GEMMs with additional modular reduction computations.

\subsubsection{Alternative ring packing methods}
\label{sec:unify:vs-others}

Unlike prior ring packing methods, ExpPack avoids both sequential loops that limit parallelism and heavy precomputation that incurs additional memory overhead.
While our unified model does not preclude alternative packing methods used in studies such as HintlessPIR~\cite{crypto-2024-hintless}, YPIR~\cite{usenixsec-2024-ypir}, and InsPIRe~\cite{iacr-2025-inspire}, many exhibit inherent limitations on GPUs.
In particular, HintlessPIR and InsPIRe rely on long $\mathcal{O}(n)$ sequential loops arising from data-dependent, stateful transformations (e.g., iterative rotations and accumulations), where each step depends on the previous, restricting parallelism.
They also rely on extensive precomputation, whose results occupy a large memory footprint; in our 64GiB \DB evaluation, InsPIRe’s CPU implementation exhausted memory even with 256GiB of DRAM.
YPIR avoids such sequential structures and exhibits computation patterns closer to ExpPack, but does not provide the same protocol-level advantages as in \S\ref{sec:unify:ExpPack:multi}.


\subsection{Putting it all together for our hybrid protocol}

We walk through the hybrid protocol with our optimizations for a 64GiB \DB organized as a $N\times 2 \times \cdots \times 2 \times D_0 = N \times D_r \times D_0 = 2^{12}\times2^{8}\times2^{16}$ structure of 1B records.
In the offline phase, the server precomputes the hint by multiplying \DB with a $D_0\times n$ random matrix, producing a $(N\times D_r)\times n$ result.
It then performs NTTs along the $N$-dimension to obtain $\mathbf{H}$, a $D_r\times n$ matrix of polynomials.
Each polynomial contains $3\times N$ INT32 elements ($\text{\#RNS}=3$), occupying 48KiB.
Thus, $\mathbf{H}$ occupies 15GiB of memory space, becoming the second largest data object in the protocol after \DB.
The client also sends reusable keys (6.09MiB) during the offline phase.

During the online phase, the client sends a length-$D_0$ INT32 vector (256KiB) for scalar-HE folding (\circled{1}) and one RLWE ciphertext (96KiB) to be expanded.
Then the online server-side computation is performed as follows:
\begin{itemize}[leftmargin=0.6cm]
    \item[\circled{1}] (\DB GEMM) INT8-INT32 GEMM between the query vector and \DB (64GiB) produces an $N\times D_r$ INT32 matrix.
    \item[\circled{3}] (ExpPack) First, expansion produces a “special” $\mathbf{C}$ ($n\times2$ polynomials, 120MiB) for ExpPack as well as RGSW ciphertexts for the remaining poly-HE folding steps. NTT is applied to the \circled{1} output, producing $D_r$ polynomials ($\mathbf{G}$, 12MiB).
    Then, INT32 GEMMs are performed to compute $\mathbf{H}\cdot\mathbf{C}+(\mathbf{0}\mid\mathbf{G})$ for Tiptoe-style packing, yielding $D_r$ output RLWE ciphertexts (24MiB).
    \item[\circled{7}] (The rest) RGSW-RLWE external products for poly-HE folding are performed $\log_2 D_r=8$ times to extract a single RLWE ciphertext (96KiB). After \circled{3}, the data size is small enough that these steps contribute little to the overall cost. The final modulus reduction~\cite{focs-2011-modulus-reduction} yields a 32KiB response (one RLWE ciphertext with two polynomials and $\text{\#RNS}=1$).
\end{itemize}

\section{\NAME: A GPU-Optimized PIR Framework}

We develop \textbf{\NAME}, a versatile GPU-optimized PIR framework based on our unified model.
While it supports a range of PIR protocols, our goal is not to cover all existing variants, but to optimize those well-suited for GPU execution.
By co-designing the protocol with GPU kernels and execution strategies, we bring PIR performance to practical levels even for large databases.

The two primary computations in PIR are GEMM and NTT.
Two types of GEMMs appear in the major execution paths of our unified model:
\begin{itemize}[leftmargin=*]
\item \textbf{INT8-INT32 plain GEMM.} This is used for scalar-HE folding (\circled{1}). As we perform modulo-$2^{32}$ operations as in SimplePIR, modular reduction effectively becomes free.
    \item \textbf{INT32 modular GEMM.} This is used for poly-HE folding (\circled{6}) and ring packing (\circled{3}). In this case, we operate on polynomials, resulting in ($\text{\#RNS} \cdot N$)-parallel GEMMs enabled by RNS decomposition and NTT. We need to perform modulo-$q_i$ operations.
\end{itemize}
%

\subsection{Scalar-HE folding with tensor-core GEMMs}

Since scalar plaintexts fit within INT8, we leverage NVIDIA GPU tensor cores~\cite{ieee-micro-2018-tensorcore} to exploit their high computational throughput.
For example, for H100 NVL GPUs, which are used in our evaluation, GEMMs with regular INT32 operations are limited to a peak throughput of 30TOPS, whereas tensor cores enable up to 1670TOPS for INT8 operations.

As tensor cores support INT8 GEMMs with INT32 accumulation, we decompose the INT32 matrix into four INT8 matrices.
For an INT8 matrix $\mathbf{A}$ and an INT32 matrix $\mathbf{B}$, we decompose $\mathbf{B}$ into INT8 matrices $\mathbf{B}_0$, $\mathbf{B}_1$, $\mathbf{B}_2$, $\mathbf{B}_3$ to compute $\mathbf{A}\cdot\mathbf{B}$ as follows:
\[
\mathbf{A} \cdot \mathbf{B_0} + 2^8\cdot(\mathbf{A} \cdot \mathbf{B_1}) + 2^{16}\cdot(\mathbf{A} \cdot \mathbf{B_2}) + 2^{24}\cdot(\mathbf{A} \cdot \mathbf{B_3}) \bmod 2^{32},
\]
where each $\mathbf{A}\cdot \mathbf{B}_i$ is computed using tensor cores, multiplications by $\{2^8, 2^{16}, 2^{24}\}$ are implemented via bit shifts, and modulo-$2^{32}$ reduction is free.

We exploit NVIDIA’s \texttt{prmt} instruction~\cite{nvidia-ptx-isa}, originally introduced for video processing, to efficiently perform the matrix decomposition.
This instruction selects four arbitrary bytes from two 32-bit registers and assembles them into a single 32-bit destination register.
Given four contiguous INT32 elements from $\mathbf{B}$, we use two \texttt{prmt} instructions to extract one byte from each element and pack them into a single 32-bit register for subsequent tensor-core operations.
This enables the decomposition to be implemented using only $2 \times (\text{\# of elements from } \mathbf{B})$ \texttt{prmt} instructions.

All these operations, including \texttt{prmt}-based decomposition, INT8-INT8 tensor-core computation, and final shift-based accumulation, are fused into a single GPU kernel to minimize memory overhead.
We further leverage advanced GPU features provided in recent NVIDIA GPUs, such as asynchronous global-to-shared memory copies, vectorized load/store instructions, matrix-specific shared-memory loads (\texttt{ldmatrix}), and swizzling~\cite{cutlass-blog}.

Our parameter and protocol choices are tailored to enable efficient tensor-core execution.
While we have taken INT8 plaintexts and INT32 ciphertexts for granted thus far, this requires setting the plaintext modulus to $2^8$ (unlike SimplePIR’s 247--991 range) and the ciphertext modulus to $2^{32}$ for scalar-HE.
In contrast, packing methods in YPIR~\cite{usenixsec-2024-ypir} and InsPIRe~\cite{iacr-2025-inspire} do not support power-of-two ciphertext moduli, requiring modulo-$q_i$ ($q_i<2^{32}$) arithmetic as in poly-HE, which introduces significant overheads in accumulation and modular reduction under tensor-core execution~\cite{hpca-2023-tensorfhe}.

\subsection{Optimizing parallel modular GEMMs}

We optimize parallel INT32 modular GEMMs used in ring packing and poly-HE folding by (1) reconciling the layout mismatch between NTT and GEMM, and (2) reducing modular reduction overhead.

In ExpPack, NTTs~\cite{hpca-2026-ive} dominate the computation of expansion and favor an $N$-innermost layout to enable contiguous accesses within each polynomial.
However, this layout is suboptimal for ($\text{\#RNS}\cdot N$)-parallel GEMMs, where threads operate on matrix tiles and require contiguous accesses within the tiles for memory coalescing and shared-memory reuse~\cite{cutlass-blog}.
To bridge this mismatch, we insert layout transpositions before and after the GEMMs to convert between polynomial-major (NTT-friendly) and tile-major (GEMM-friendly) layouts.
This enables coalesced global memory accesses, higher shared-memory reuse, and larger GEMM tile usage, improving arithmetic intensity and throughput.

%
We further reduce modular reduction overhead via lazy reduction~\cite{asplos-2026-cheddar, tches-2021-100x}.
Using RNS primes smaller than $2^{29}$ and 64-bit accumulators, we defer reduction for up to 32 multiply-accumulates without overflow, amortizing reduction costs and lowering instruction overhead.


\subsection{Number-theoretic transform (NTT)}

We optimize NTT execution by employing a single-kernel design with Montgomery reduction.
For the NTT with $N=2^{12}$, whose working set fits within on-chip resources, we can implement it with a single kernel, eliminating inter-kernel synchronization overhead~\cite{tjs-2022-gpu-ntt, cal-2024-pbs}.
We also adopt Montgomery reduction~\cite{1985-montgomery} instead of Barrett reduction~\cite{eurocrypt-1986-barrett}, as it is better suited for the 32-bit integer datapath of GPUs with fewer instructions~\cite{asplos-2026-cheddar}.

We also fuse kernels to reduce memory traffic between the NTT and its adjacent kernels.
Specifically, we fuse the initial RNS decomposition with the NTT in ExpPack (\S\ref{sec:unify:ExpPack}), reducing the DRAM transfers for temporary results.
We also fuse modulus reduction~\cite{focs-2011-modulus-reduction} with inverse NTT.



\subsection{Memory-capacity-aware stream-wise buffering}

Although we avoid \DB bloating, certain intermediate data objects occupy large memory space.
In particular, the second folding in our double-packing protocol multiplies a compressed intermediate ($8N\times D_1$ INT8 per client) with scalar-HE ciphertexts ($D_1\times(n+1)$ INT32), producing an $8N\times(n+1)$ result (160MiB) per client, totaling 10GiB for 64 batches; subsequent NTTs for ring packing further triple this footprint.

To balance batching parallelism and memory efficiency, we introduce stream-wise buffering.
For $B$-batch execution, we partition execution across CUDA streams, allocate a fixed per-stream buffer sized for a single batch, and assign each stream $B/\text{\#stream}$ batches.
Each stream processes its batches sequentially, reusing the buffer, reducing intermediate memory usage from $\mathcal{O}(B)$ to $\mathcal{O}(\text{\#stream})$ while preserving parallelism.
We apply this selectively to memory-intensive stages.

\subsection{Multi-GPU support}

To further scale throughput and support larger \DB sizes, we extend \NAME with multi-GPU support.
We devise two parallelization strategies for distributing data and computation, both employed within a single protocol.

\begin{figure}[t!]
    \centering
    \includegraphics[width=0.99\linewidth]{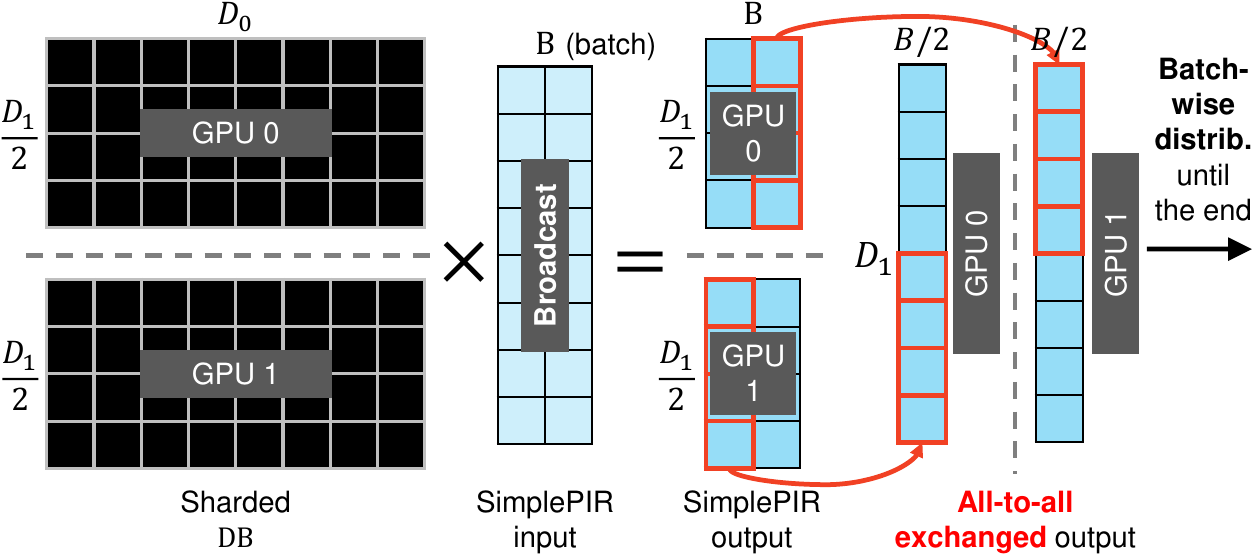}
    \Description{}
    \caption{Two-GPU online computation flow for SimplePIR.}
    \label{fig:multi-gpu}
\end{figure}

First, because \DB dominates DRAM usage, we shard it across GPUs to support \DB sizes exceeding a single GPU’s memory capacity.
Figure~\ref{fig:multi-gpu} shows an example for SimplePIR, where each GPU stores a $(D_1/\text{\#GPU}) \times D_0$ partition of \DB and independently performs GEMMs with the input queries (a $D_0 \times B$ matrix), which are broadcast to all GPUs.

After the first folding, we perform inter-GPU communications to transition to a batch-wise distribution, eliminating the need for further communications in later stages.
If the first folding produces a $(D_1/\text{\#GPU}) \times B$ output, we perform an all-to-all exchange so that each GPU holds a $D_1 \times (B/\text{\#GPU})$ partition of the aggregated result, effectively reducing the per-GPU batch size to $B/\text{\#GPU}$.
As the SimplePIR hint is shared across all queries and clients, it must reside on every GPU for further PIR steps.
Thus, we instead perform an all-gather during hint precomputation.

\setlength{\tabcolsep}{2pt}
\begin{table*}[t]
    \centering
    \caption{Performance comparison of hintless PIR implementations across various \DB sizes. The first three use only poly-HE folding; the next six use scalar-HE folding based on SimplePIR (S) or DoublePIR (D)~\cite{usenixsec-2023-simplepir}; and the last two combine both poly-HE and scalar-HE folding. Green cells denote the best value in each group. We report single-batch ($B\!=\!1$) server-side online execution times in milliseconds. Communication per query (Comm./query) includes non-reusable keys, but excludes reusable keys (Reuse-keys). We used one H100 NVL 94GB GPU for \NAME. OoM refers to out-of-memory failures.}
    \label{tab:eval:vs_cpu}
    \vspace{-0.05in}
    {
\small
    \begin{tabularx}{0.99\linewidth}{c|l|CCC|CCCCCC|CC}
    \toprule
    \multirow{3}{*}{\begin{tabular}{c}\DB \\ size\end{tabular}} & System & CPU & CPU & \textbf{\NAME} & CPU & CPU & CPU & \textbf{\NAME} & \textbf{\NAME} & \textbf{\NAME} & CPU & \textbf{\NAME}\\
     & \multirow{2}{*}{Protocol} & \multirow{2}{*}{Spiral} & Onion & Onion & Hintless & YPIR & YPIR & ExpPack & ExpPack & Double-  & \multirow{2}{*}{InsPIRe} & \multirow{2}{*}{Hybrid}\\
     & & & PIRv2 & PIRv2 & PIR & (S) & (D) & (S) & (D) & packing & & \\
    \midrule
    & Retrieved records & 16KiB & 16KiB & 16KiB & 1B & 32KiB & 1B & 32KiB & 1B & 4KiB & 4KiB & 4KiB\\ 
    \multirow{3}{*}{\begin{tabular}{c}1\\GiB\end{tabular}} & \DB storage & \cellcolor{red!25}8GiB & \cellcolor{red!25}4GiB & \cellcolor{red!25}4GiB & 1GiB & 1GiB & 1GiB & 1GiB & 1GiB & 1GiB & 1GiB & 1GiB \\
   & Exec. time ($B\!=\!1$) & 3,310 & 1,160 & 4.1\cellcolor{green!25} & 647 & 355 & 134 & 3.1\cellcolor{green!25} & 10.7 & 9.8 & 4,658 & 9.6\cellcolor{green!25}\\
    & Comm./query & 96KiB\cellcolor{green!25} & 128KiB & 256KiB & 1.94MiB & 924KiB & 924KiB & 480KiB & 395KiB & 357KiB\cellcolor{green!25} & 271KiB & 132KiB\cellcolor{green!25}\\
    & Reuse-keys/client & 16.0MiB & 4.00MiB\cellcolor{green!25} & 8.00MiB & 384KiB & 0\cellcolor{green!25} & 0\cellcolor{green!25} & 5.16MiB & 5.16MiB & 5.16MiB & 0\cellcolor{green!25} & 6.09MiB\\
    \midrule
    & Retrieved records & 16KiB & 16KiB & 16KiB & 1B & 64KiB & 1B & 64KiB & 1B & 4KiB & 4KiB & 4KiB\\ 
    \multirow{3}{*}{\begin{tabular}{c}4\\GiB\end{tabular}} & \DB storage & \cellcolor{red!25}32GiB & \cellcolor{red!25}16GiB & \cellcolor{red!25}16GiB & 4GiB & 4GiB & 4GiB & 4GiB & 4GiB & 4GiB & 4GiB & 4GiB \\
     & Exec. time ($B\!=\!1$) & 10,150 & 3,896 & 10.5 \cellcolor{green!25} & 1,242 & 883 & 430 & 4.4\cellcolor{green!25} & 19.3 & 10.8 & 6,919 & 10.6\cellcolor{green!25}\\
    & Comm./query & 96KiB\cellcolor{green!25} & 128KiB & 256KiB & 3.69MiB & 1.28MiB & 1.28MiB & 864KiB & 651KiB & 369KiB\cellcolor{green!25} & 292KiB & 144KiB\cellcolor{green!25}\\
    & Reuse-keys/client & 16.0MiB & 4.00MiB\cellcolor{green!25} & 8.00MiB & 384KiB & 0\cellcolor{green!25} & 0\cellcolor{green!25} & 5.16MiB & 5.16MiB & 5.16MiB & 0\cellcolor{green!25} & 6.09MiB\\
    \midrule
    & Retrieved records & 16KiB & 16KiB & 16KiB & 1B & 128KiB & 1B & 128KiB & 1B & 4KiB & 4KiB & 4KiB\\ 
    \multirow{3}{*}{\begin{tabular}{c}16\\GiB\end{tabular}} & \DB storage & \cellcolor{red!25}128GiB & \cellcolor{red!25}64GiB & \cellcolor{red!25}64GiB & 16GiB & 16GiB & 16GiB & 16GiB & 16GiB & 16GiB & 16GiB & 16GiB \\
    & Exec. time ($B\!=\!1$) & 51,960 & 15,144 & 35.9 \cellcolor{green!25} & 3,183 & 2,654 & 1,680 &9.3\cellcolor{green!25} & 38.6 & 15.3 & 12,738 & 15.1\cellcolor{green!25}\\
    & Comm./query & 96KiB\cellcolor{green!25} & 128KiB & 256KiB & 7.19MiB & 2.03MiB & 2.03MiB & 1.59MiB & 1.14MiB & 417KiB\cellcolor{green!25} & 376KiB & 192KiB\cellcolor{green!25}\\
    & Reuse-keys/client & 16.0MiB & 4.00MiB\cellcolor{green!25} & 8.00MiB & 384KiB & 0\cellcolor{green!25} & 0\cellcolor{green!25} & 5.16MiB & 5.16MiB & 5.16MiB & 0\cellcolor{green!25} & 6.09MiB\\
    \midrule
    & Retrieved records & 16KiB & 16KiB & 16KiB & 1B & 256KiB & 1B & 256KiB & 1B & 4KiB & 4KiB  & 4KiB\\ 
    \multirow{3}{*}{\begin{tabular}{c}64\\GiB\end{tabular}} & \DB storage & \cellcolor{red!25}512GiB & \cellcolor{red!25}256GiB & \cellcolor{red!25}256GiB & 64GiB & 64GiB & 64GiB & 64GiB & 64GiB & 64GiB & 64GiB & 64GiB  \\
    & Exec. time ($B\!=\!1$) & \cellcolor{gray!20} OoM& \cellcolor{gray!20} OoM & \cellcolor{gray!20} OoM & 10,637 & 9,883 & 7,076 &28.3\cellcolor{green!25} & 90.8 & 33.0 & \cellcolor{gray!20} OoM & 33.5\cellcolor{green!25}\\
    & Comm./query & 96KiB\cellcolor{green!25} & 128KiB & 256KiB & 14.2MiB & 3.53MiB & 3.53MiB & 3.09MiB & 2.14MiB & 609KiB\cellcolor{green!25} & 712KiB & 384KiB\cellcolor{green!25}\\
    & Reuse-keys/client & 16.0MiB & 4.00MiB\cellcolor{green!25} & 8.00MiB & 384KiB & 0\cellcolor{green!25} & 0\cellcolor{green!25} & 5.16MiB & 5.16MiB & 5.16MiB & 0\cellcolor{green!25} & 6.09MiB\\
    \bottomrule    
    \end{tabularx}
    }
\end{table*}
\setlength{\tabcolsep}{6pt}

\section{Implementation}
\label{sec:impl}

\NAME is implemented in C++ and CUDA.
We implement the INT8-INT32 GEMM using the CuTE layout abstractions from CUTLASS~\cite{github-cutlass}, and leverage NVIDIA Collective Communications Library (NCCL)~\cite{nvidia-nccl} for efficient inter-GPU communications.
We use separate CUDA-core-based kernels for the INT8-INT32 matrix-vector multiplications in single-batch execution as they are memory-bound and cannot exploit tensor-core throughput.
\NAME focuses on accelerating server-side PIR operations, the primary performance bottleneck; client-side operations are implemented for evaluation purposes using SEAL~\cite{fcds-2017-seal}, a CPU-based HE library.

We follow standard practices~\cite{2021-standard, sosp-2023-tiptoe, usenixsec-2023-simplepir, ccs-2021-onionpir, iacr-2025-onionpirv2} for selecting the parameters.
For scalar-HE, we set the plaintext modulus to $2^8$ and the ciphertext modulus to $2^{32}$, which is reduced to $2^{18}$ after each scalar-HE folding step.
For poly-HE (RLWE and RGSW), we set the plaintext modulus to $2^{18}$, and the ciphertext modulus to $q \simeq 2^{87}$ ($\text{\#RNS}=3$), which can be reduced to $q\simeq2^{29}$ ($\text{\#RNS}=1$) through modulus reduction~\cite{focs-2011-modulus-reduction}, where slightly larger parameters are used for implementing OnionPIRv2~\cite{iacr-2025-onionpirv2}.
We verify that these parameters achieve 128-bit security when using degrees of $n=1{,}280$ and $N=2^{12}$ based on Lattice Estimator~\cite{jmc-2025-lattice-estimator}.
Parameter configurations are more detailed in Appendix~\ref{app:symbol} and Appendix~\ref{app:security}.

\section{Evaluation}
\label{sec:eval}

\subsection{Experimental setup}

We tested \NAME across multiple PIR protocols, including existing ones---OnionPIRv2~\cite{iacr-2025-onionpirv2} and SimplePIR/DoublePIR~\cite{usenixsec-2023-simplepir}; hintless protocols differing from prior work~\cite{usenixsec-2024-ypir, sosp-2023-tiptoe, crypto-2024-hintless} only in ring packing---ExpPack(S) (SimplePIR-based: \circled{1} $\rightarrow$ \circled{3}) and ExpPack(D) (DoublePIR-based: \circled{1} $\rightarrow$ \circled{2} $\rightarrow$ \circled{1} $\rightarrow$ \circled{3}); plus our double-packing and hybrid protocols.

We primarily evaluated the performance of \NAME on an H100 NVL 94GB GPU with an effective HBM bandwidth of 3.48TB/s.
To demonstrate multi-GPU scalability, we also conducted experiments using two H100 NVL 94GB GPUs connected via NVLink 3.0, offering a bandwidth of 530.16GB/s.
All reported execution times reflect server-side online execution times only, with throughput measured in queries per second (QPS).
When \DB contains $D$ 1B records, we set $D_0\approx\sqrt{D}$ for SimplePIR and DoublePIR (both with and without ExpPack).
For double-packing and hybrid protocols, we used $D/D_0=2^{20}$ to minimize communication, except for 64GiB \DB, where we used $D/D_0=2^{19}$ by default to avoid out-of-memory failures under large-batch execution (Table~\ref{tab:eval:vs_cpu} still uses $2^{20}$ as it reports single-batch results).

We compare \NAME against state-of-the-art PIR studies, including Spiral~\cite{sp-2022-spiral}, OnionPIRv2~\cite{iacr-2025-onionpirv2},  HintlessPIR~\cite{crypto-2024-hintless}, YPIR~\cite{usenixsec-2024-ypir}, InsPIRe~\cite{iacr-2025-inspire}, and SimplePIR/DoublePIR~\cite{usenixsec-2023-simplepir}.
Their open-source CPU versions were executed on a system with an Intel Xeon Gold 6348 CPU and 256GiB of DRAM.
Configurations for the baselines are described in Appendix~\ref{app:experimental}.

\subsection{\NAME vs. CPU baselines}

Our extensive comparison against prior CPU baselines (Table~\ref{tab:eval:vs_cpu}) shows that \NAME achieves orders-of-magnitude higher performance in server-side PIR computations than all the baselines.
Compared to DoublePIR-based YPIR~\cite{usenixsec-2024-ypir}, which shows the best performance among the baselines, even our slowest configuration, ExpPack(D) demonstrates 12.5--77.9$\times$ improvements.
Notably, the performance gains increase with the \DB size---12.5$\times$ for 1GiB and 77.9$\times$ for 64GiB---highlighting \NAME's strong scalability to large databases.

Owing to our memory-capacity-aware design, which introduces minimal overhead for auxiliary data objects, \NAME can efficiently accommodate large databases within limited memory budgets.
In contrast, InsPIRe~\cite{iacr-2025-inspire} and poly-HE folding approaches, such as Spiral~\cite{sp-2022-spiral} and OnionPIRv2~\cite{iacr-2025-onionpirv2}, encounter out-of-memory (OoM) failures even with 256GiB of host memory.
\NAME, however, successfully handles up to 64GiB of \DB entirely on the GPU, despite having only 94GiB of available memory.

Our new protocol constructions demonstrate promising reductions in communication cost while maintaining competitive server-side computation performance, even with their additional computation overhead.
The double-packing and hybrid protocols achieve the lowest per-query communication cost within their respective protocol groups.
At the same time, their server-side performance matches that of simple designs such as ExpPack(S) for large databases.
For 64GiB \DB, ExpPack(S) requires 28.3ms of server computation, whereas the double-packing and hybrid protocols require comparable 33.0ms and 33.5ms, respectively.

64GiB PIR execution times of 33.0--33.5ms with our protocols are only 1.67--1.69$\times$ above the hard latency bound imposed by \DB access time.
Simply streaming 64GiB of data from DRAM requires 19.7ms on our evaluated GPU system.

\NAME also excels in executing existing protocols.
Compared to the CPU implementation of OnionPIRv2~\cite{iacr-2025-onionpirv2}, \NAME achieves 283--422$\times$ performance improvements with the identical protocol.
Although \NAME currently incurs twice the communication cost of the CPU baseline, this can be equalized by replacing half of the transmitted data with random seeds~\cite{security-2021-mulpir}, as done in the baseline.

\setlength{\tabcolsep}{4pt}
\begin{table}[t]
    \centering
    \caption{Throughput (QPS) for plain SimplePIR/DoublePIR based on the CPU implementation of YPIR~\cite{usenixsec-2024-ypir}, which does not support batching, and 32-batched \NAME.}
    \label{tab:eval_simple}
    \vspace{-0.05in}
{
\small
    \begin{tabular}{c|ccc|ccc}
    \toprule
    \DB & \multicolumn{3}{c|}{SimplePIR} &  \multicolumn{3}{c}{DoublePIR}\\  
    size & CPU & \NAME & Improv. & CPU & \NAME & Improv.\\
    \midrule
    1GiB & 11.36 & 8,244 & \phantom{0,}756$\times$ & 9.80 & 1,838 & 188$\times$ \\
    4GiB & \phantom{0}2.54 & 2,201 & \phantom{0,}867$\times$ & 2.27 & \phantom{0,}779 & 343$\times$ \\
    16GiB & \phantom{0}0.64 &  \phantom{0,}565 & \phantom{0,}883$\times$ & 0.60 &   \phantom{0,}294 & 490$\times$ \\
    64GiB & \phantom{0}0.14 & \phantom{0,}144 & 1,029$\times$ & 0.14 & \phantom{0,0}98 & 700$\times$ \\
    \bottomrule
    \end{tabular}
}
\end{table}
\setlength{\tabcolsep}{6pt}

Support for multi-client batching~\cite{hpca-2026-ive} can further widen the performance gap when evaluating throughput (QPS).
While prior CPU-based studies provide only limited batching capabilities, \NAME fully supports batching as long as memory space is sufficient.
As shown in Table~\ref{tab:eval_simple}, under identical SimplePIR/DoublePIR protocols, 32-batch execution with \NAME achieves 188--1,029$\times$ higher QPS compared to the single-batch CPU implementation~\cite{usenixsec-2024-ypir}.

\setlength{\tabcolsep}{2pt}
\begin{table}[t]
    \centering
    \caption{Throughput (QPS) across various \DB sizes for \NAME (H100 NVL 94GB) using a batch size of 32 and prior GPU-based PIR studies. OoM refers to out-of-memory failures.}
    \label{tab:qps_comparison}
{
    \small
    \begin{tabular}{c|cc|ccccc}
\toprule
    \DB &\multicolumn{2}{c|}{Prior work} &\multicolumn{5}{c}{\NAME}\\
    size & PIRon & Shift & Onion & ExpPack & ExpPack & Double- &\multirow{2}{*}{Hybrid}  \\
    (GiB) & GPU & PIR\textsuperscript{$\dagger$} & PIRv2 & (S) & (D) & packing\\
    \midrule
    1 & 7.94 & 12.50 & 516.6 & 480.6 & 461.5 & 190.4 & 244.9 \\
    4 & 2.37 & \phantom{0}3.61 & 177.8 & 465.4 & 405.0 & 186.4  & 236.1 \\
    16 & OoM & \phantom{0}0.98 & \phantom{0}48.6 & 397.2 & 303.3 & 172.3 & 219.6 \\
    64 & OoM & \phantom{0}0.25 & OoM & 232.8 & 168.3 & 163.1 & 207.4 \\
    \bottomrule
    \end{tabular}
}
    {
    \small
    \begin{itemize}[leftmargin=*]
    \item[$\dagger$] Reported values from the paper~\cite{ccs-2025-shiftpir} using an RTX 4090.
    \end{itemize}
    }
\end{table}
\setlength{\tabcolsep}{6pt}

\begin{figure*}[t!]
    \centering
    \includegraphics[width=0.99\linewidth]{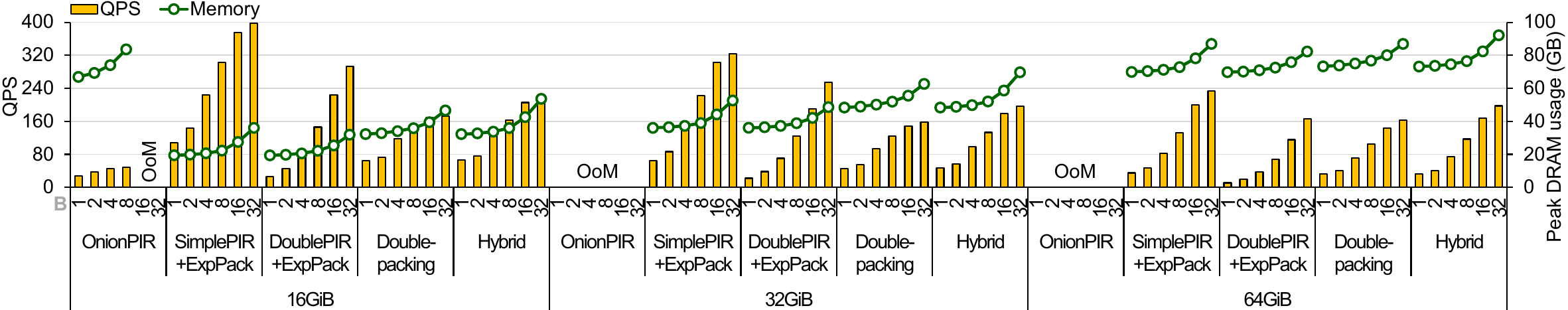}
    \Description{}
    \caption{Queries per second (QPS) and peak DRAM memory usage for different \DB sizes, PIR protocols, and batch sizes ranging from 1 to 32. We used one H100 NVL 94GB GPU with \NAME. OoM refers to out-of-memory failures.}
    \label{fig:batch_sensitivity}
\end{figure*}

\subsection{\NAME vs. GPU baselines}

Several GPU-based PIR studies also exist.
We compare \NAME against PIRonGPU~\cite{github-PIRonGPU}, which was executed on the same GPU system, as well as the reported performance of ShiftPIR~\cite{ccs-2025-shiftpir}.
PIRonGPU and ShiftPIR belong to poly-HE folding approaches, which incur significant \DB bloating due to NTT, limiting scalability to large \DB sizes.
ShiftPIR~\cite{ccs-2025-shiftpir} attempts to mitigate this by partitioning \DB and overlapping host-to-device transfers with GPU computation, but becomes constrained by the low host-device bandwidth.

As shown in Table~\ref{tab:qps_comparison}, \NAME consistently outperforms prior GPU-based PIR implementations across all \DB sizes and all protocol configurations.
Even when using OnionPIRv2, which also incurs \DB bloating, \NAME achieves 65.1--75.0$\times$ and 41.3--49.6$\times$ QPS improvements compared to PIRonGPU and ShiftPIR, respectively.
Furthermore, using our protocols without \DB bloating (all protocols in Table~\ref{tab:qps_comparison} except OnionPIRv2) yields larger QPS gains, reaching 175.7--405.1$\times$ for 16GiB \DB and 652.6--931.3$\times$ for 64GiB.

\subsection{Sensitivity study: Execution parameters}

Figure~\ref{fig:batch_sensitivity} shows how throughput (QPS) varies with \DB size, batch size, and protocol.
For all protocols, QPS increases steadily as the batch size grows.
Relative to single-batch execution, 32-batch execution improves GPU throughput by 3.69--6.62$\times$, 11.33--15.06$\times$, 2.64--4.95$\times$, and 3.33--6.02$\times$ for ExpPack(S), ExpPack(D), double-packing, and hybrid protocols, respectively.
The particularly strong scaling of ExpPack(D) stems from its minimal reliance on ring packing, which is applied only to $\mathcal{O}(n^2)$ data elements and is thus independent of the \DB dimensions.
This implies that our tensor-core GEMM implementation for dimension folding scales well with batching.

The QPS gap among the protocols narrows as the \DB size increases, which makes our double-packing and hybrid protocols more attractive for large databases.
Using 32-batch execution as the reference point, for a 16GiB \DB, ExpPack(S) achieves 2.31$\times$ and 1.81$\times$ higher QPS than the double-packing and hybrid protocols, respectively.
When the \DB size increases to 64GiB, however, these gaps shrink to 1.43$\times$ and 1.18$\times$.
This is because the overall execution time becomes increasingly dominated by the \DB GEMM performed in the first scalar-HE folding step, while steps such as expansion remain a constant overhead.

Figure~\ref{fig:batch_sensitivity} also shows the peak device memory usage for each case.
For OnionPIR, even a single-batch execution with a 16GiB \DB already uses 66.9GiB of memory because of the 4$\times$ \DB bloating.
The other protocols based on scalar-HE folding do not suffer from this issue; even for a 64GiB \DB and a batch size of up to 32, they use only 69.8--92.1GiB of memory.

\begin{figure}
    \centering
    \includegraphics[width=0.99\linewidth]{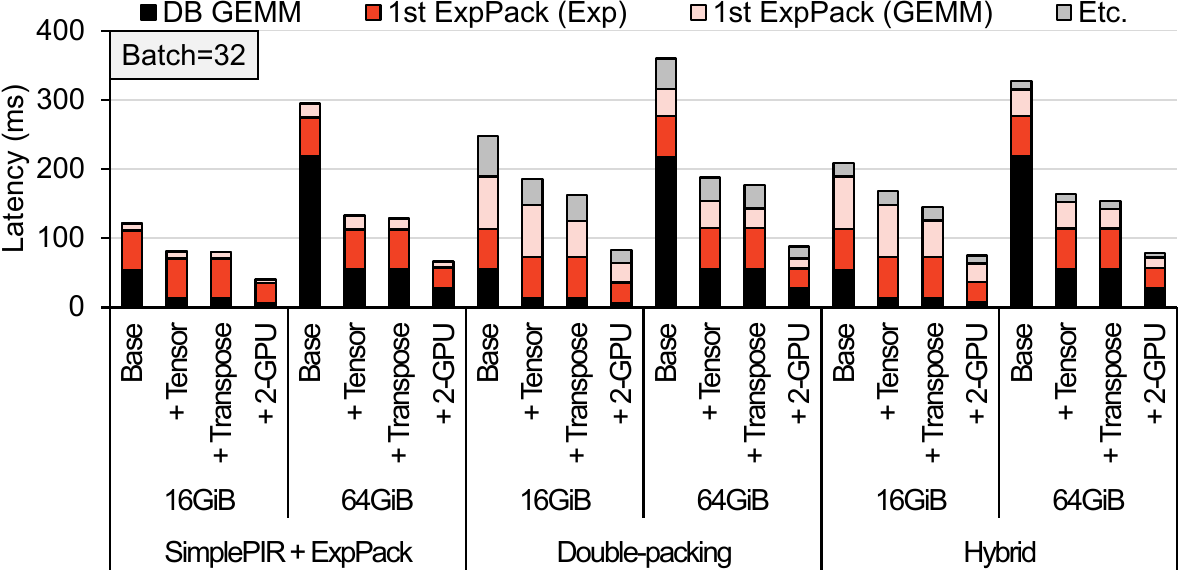}
    \Description{}
    \caption{Online server computation latency as our GPU optimizations---tensor-core INT8-INT32 GEMMs (+Tensor), transposed INT32 GEMMs (+Transpose), and two-GPU execution (+2-GPU)---are incrementally applied. We utilized H100 NVL 94GB GPUs with NVLink connections.}
    \label{fig:gpu_opt_ablation}
\end{figure}

\subsection{Sensitivity study: GPU optimizations}

Figure~\ref{fig:gpu_opt_ablation} shows the 32-batch execution time breakdown as we incrementally apply our GPU optimizations targeting GEMMs.
For the 64GiB \DB, without tensor cores (Base), the initial \DB GEMM uses standard INT32 operations and accounts for 60.5--74.0\% of total server-side online computation time.
Our tensor-core \DB GEMM implementation accelerates this by 3.89--4.05$\times$, reducing its share to 29.8--42.2\%.
Applying transposed INT32 GEMMs further accelerates the GEMMs in the first ExpPack computation by 1.09--1.44$\times$.
Together, these two GEMM optimizations deliver 1.44--1.53$\times$ end-to-end speedups for the 16GiB \DB, and much larger 2.04--2.29$\times$ speedups for the 64GiB \DB.

Additionally, thanks to our efficient multi-GPU partitioning strategies, \NAME
achieves near-perfect scaling with 1.94--2.01$\times$ speedups when using two GPUs,
where speedups also come from reduced per-GPU working set.
While \NAME supports more GPUs, we leave their evaluation as future work.

\subsection{Deeper analysis of the hybrid protocol}

\begin{figure}
    \centering
    \includegraphics[width=0.99\linewidth]{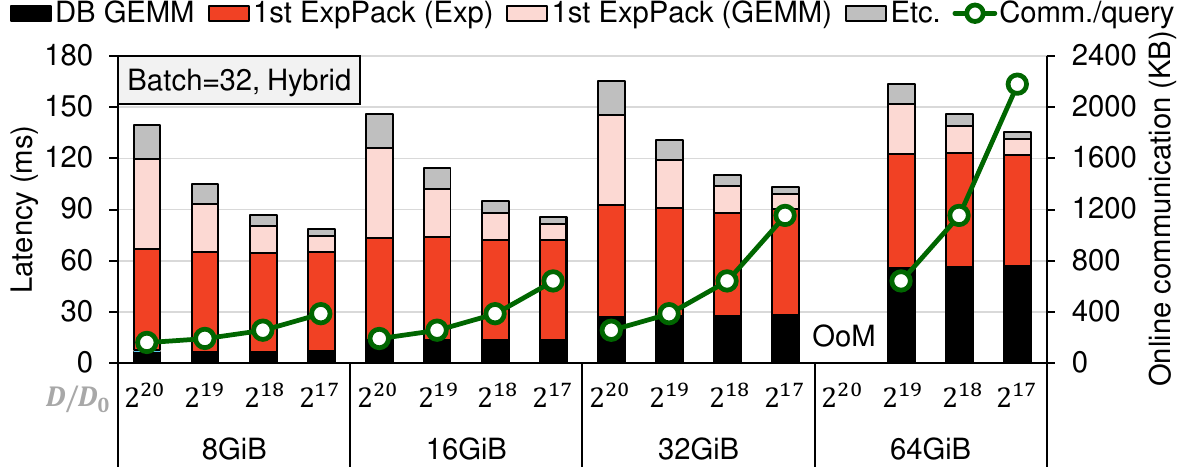}
    \Description{}
    \caption{Latency breakdown and online communication as $D / D_0$ varies in the hybrid protocol on an H100 NVL 94GB GPU. OoM refers to out-of-memory failures.}
    \label{fig:hybrid_sensitivity}
\end{figure}

Determining the \DB dimensionality is a critical design choice for protocol constructions, with particularly substantial impacts on the hybrid protocol.
Suppose \DB contains $D$ records in total.  
Figure~\ref{fig:hybrid_sensitivity} shows that varying $D_0$ creates a trade-off between server computation and client-server communication.  
Increasing $D_0$ (decreasing $D/D_0$) yields more compressed outputs (compression factor: $\frac{4(n+1)}{D_0}\times$) from the initial folding, reducing the data volume for ring packing.
In contrast, $D_0$ has negligible impact on \DB GEMM time, as the $(D/D_0) \times B \times D_0$ GEMM involves the same $D \cdot B$ multiply-accumulates regardless of $D_0$, for batch size $B$.
Therefore, increasing $D_0$ reduces overall server execution time but increases communication volume ($\mathcal{O}(D_0)$) to send the encrypted one-hot representation of the index in the $D_0$ dimension.

We primarily used $D/D_0=2^{20}$ in our experiments to minimize server-client communication, as computational costs are largely mitigated by our optimizations.
If we instead opt for more communication-heavy settings with $D/D_0 = 2^{17}$, we can obtain 1.59--1.76$\times$ speedups (vs. $2^{20}$) for \DB sizes smaller than 64GiB, and 1.21$\times$ for 64GiB (vs. $2^{19}$).
We leverage this trade-off to show multiple data points in Figure~\ref{fig:intro}.

\section{Related Work}
\label{sec:related}

Since the seminal work of Chor et al.~\cite{jacm-1998-chorpir}, numerous PIR protocols have been proposed.
XPIR~\cite{popets-2016-xpir} employs a single poly-HE folding step.
SealPIR~\cite{sp-2018-sealpir} introduces query expansion and performs multiple poly-HE folding steps with a reinterpretation step~\cite{crypto-2012-bfv, iacr-2012-bfv2}.
WhisPIR~\cite{iacr-2024-whispir} later introduces an expansion method with reduced expansion key overhead.
Building on SealPIR, MulPIR~\cite{security-2021-mulpir} employs RLWE inter-ciphertext multiplications, while OnionPIR~\cite{ccs-2021-onionpir} further improves efficiency by using RGSW-RLWE external products~\cite{jc-2020-tfhe} for non-initial folding steps.
OnionPIRv2~\cite{iacr-2025-onionpirv2} further optimizes the design with hardware-friendly implementations and protocol fine-tuning.
Spiral~\cite{sp-2022-spiral} explores MLWE~\cite{dcc-2015-mlwe} in place of RLWE, and Respire~\cite{ccs-2024-respire} leverages subring-based RLWE dimension reduction for small-record databases.
KsPIR~\cite{ccs-2024-kspir} instead proposes using encrypted RLWE matrix-vector multiplications for poly-HE folding.

In parallel, scalar-HE dimension folding has gained traction since its introduction in SimplePIR~\cite{usenixsec-2023-simplepir}.
SimplePIR reduces communication by requiring clients to download a large reusable hint.
However, this hint must be replaced for every \DB update and requires a large client-side storage.
Hintless protocols, including
Tiptoe~\cite{sosp-2023-tiptoe}, HintlessPIR~\cite{crypto-2024-hintless}, and YPIR~\cite{usenixsec-2024-ypir}, apply various ring packing methods to SimplePIR, building on packing techniques used in other contexts~\cite{acns-2021-cdks, asia-2017-column-ring-pack, sp-2021-pegasus}.
There are potentially more packing methods that may be applicable to PIR, such as HERMES~\cite{crypto-2023-hermes}.
InsPIRe~\cite{iacr-2025-inspire} proposes a new packing method that leverages heavy precomputation to reduce online computation cost, and introduces a poly-HE folding technique that interpolates \DB entries into a function.

Despite extensive work on improving PIR protocols, relatively few studies focus on their hardware acceleration.
PIRonGPU~\cite{github-PIRonGPU} provides an open-source GPU implementation of SealPIR~\cite{sp-2018-sealpir} using the HEonGPU library~\cite{iacr-2024-HEonGPU}.
ShiftPIR~\cite{ccs-2025-shiftpir} proposes a GPU-aware protocol that reduces client overhead by partially offloading query generation to the server.
Lam et al.~\cite{ASPLOS-2024-gpupir} and CIP-PIR~\cite{usenixsec-2022-gpupir} explore GPU acceleration for multi-server PIR protocols, which rely on non-collusion assumptions that we consider restrictive and thus do not address in this work.

Several recent architecture proposals also target PIR acceleration.
IVE~\cite{hpca-2026-ive} proposes an ASIC accelerator design dedicated for processing batches of OnionPIR queries.
INSPIRE~\cite{isca-2022-inspire} (distinct from InsPIRe) introduces an in-storage accelerator for PIR on SSDs.
SmartPIR~\cite{micro-2025-smartpir} extends this approach by offloading HE operations to computational storage devices (CSDs).
Conflux~\cite{hpca-2026-conflux} similarly leverages CSDs for keyword PIR, an extension of index-based PIR.
\section{Conclusion}
\label{sec:conclusion}

We presented \NAME, a versatile GPU framework that rethinks the design and implementation of PIR through a unified, GPU-centric perspective.
By introducing a unified analytical model, we expose fundamental trade-offs in prior PIR protocols, particularly the tension between ciphertext and \DB (plaintext) bloating.
Guided by these insights, we propose double-packing and hybrid protocols that improve communication efficiency without incurring prohibitive memory capacity demands.
A key enabler is ExpPack, a novel ring packing technique that reduces communication cost while exposing massive parallelism well-suited for GPUs.
\NAME also includes various architecture-aware optimizations, including tensor-core-accelerated GEMMs, optimized NTT execution, memory-efficient scheduling, and multi-GPU support.
Extensive evaluation shows that \NAME achieves orders-of-magnitude performance improvements over prior CPU- and GPU-based PIR systems, while also reducing communication costs and avoiding out-of-memory failures.

\bibliographystyle{ACM-Reference-Format}
\bibliography{refs_with_url}

\appendix
\section{Cryptographic Background \& Symbols}
\label{app:symbol}

We introduce the cryptographic constructions of LWE-based scalar-HE and RLWE-based poly-HE in this appendix.

\subsection{LWE scalar-HE and its parameters}

LWE has a plaintext modulus $p_\text{LWE}$ and a ciphertext modulus $q_\text{LWE}$.
For an LWE degree $n$, the client samples a length-$n$ secret vector $\mathbf{s}=(s_0,s_1,\cdots,s_{n-1})\in\{-1, 0, 1\}^n$ from a uniform ternary distribution, a length-$n$ random vector $\mathbf{a}=(a_0, a_1, \cdots, a_{n-1})\in\mathbb{Z}_{q_\text{LWE}}^n$ from a uniform distribution in $\mathbb{Z}_{q_\text{LWE}}$, and an error value $e\in\mathbb{Z}$ from a discrete Gaussian distribution with a standard deviation $\sigma$.
Typically, $\sigma$ is set to 3.2 and $e$ is clipped to ensure that $|e|<6\sigma$.

We can encrypt a single plaintext scalar $m\in\mathbb{Z}_{p_\text{LWE}}$ by the following:
\[
    \mathrm{Encrypt}(m)\rightarrow\mathtt{ct}=(\mathbf{a},b)\in\mathbb{Z}_{q_\text{LWE}}^{n+1},
\]
\[
    b + \mathbf{a}^T\mathbf{s} = b + \sum_{i=0}^{n-1} a_i \cdot s_i = \Delta \cdot m + e \pmod{q_\text{LWE}}
\]
Here, $\Delta$ is called the scale and is determined as $\Delta=\frac{q_\text{LWE}}{p_\text{LWE}}$.

It is convenient to define the phase of an LWE ciphertext:
\[
\phi(\mathtt{ct}) = b + \mathbf{a}^T\mathbf{s} = \Delta \cdot m + e \pmod{q_\text{LWE}},
\]
\[
\phi(\mathtt{ct})\in\left[-\frac{q_\text{LWE}}{2}, \frac{q_\text{LWE}}{2} - 1\right]
\]

The client can recover the plaintext $m\in[-\frac{p_\text{LWE}}{2}, \frac{p_\text{LWE}}{2}]$ by computing
\[
\mathrm{Decrypt}(\mathtt{ct})=\mathrm{round}\left(\frac{1}{\Delta} \cdot \phi(\mathtt{ct})\right) = \mathrm{round}(m + \frac{e}{\Delta}) \rightarrow m,
\]
provided that $|e|<\frac{\Delta}{2}$.

For ciphertexts $\mathtt{ct}=(\mathbf{a}, b)$ and $\mathtt{ct}'=(\mathbf{a}', b')$, we can perform following inter-ciphertext additions and plaintext-ciphertext multiplications:
\[
\mathtt{ct} + \mathtt{ct}' = (\mathbf{a} + \mathbf{a}', b + b') \in \mathbb{Z}_{q_\text{LWE}}^{n+1}
\]
\[
c \cdot \mathbf{ct}_i = (c \cdot \mathbf{a}, c \cdot b) \in \mathbb{Z}_{q_\text{LWE}}^{n+1}
\]
The phase of the result will be (for sufficiently small $c$ and $m$):
\[
\phi(\mathtt{ct} + \mathtt{ct}') = \Delta \cdot (m + m') + (e + e') = \phi(\mathtt{ct}) + \phi(\mathtt{ct}')  
\]
\[
\phi(c \cdot \mathbf{ct}) = \Delta \cdot (c \cdot m) + (c \cdot e) = c \cdot \phi(\mathtt{ct})
\]

In this paper, we exclusively use the following parameters for LWE: $p_\text{LWE}=2^8$, $q_\text{LWE}=2^{32}$, $\sigma=3.2$, $n=1{,}280$, and $\Delta=2^{24}$.
The notations $p_\text{LWE}$ and $q_\text{LWE}$ are only used in this subsection of appendix.

\subsection{RLWE poly-HE and its parameters}

RLWE also has a plaintext modulus $p$ and a ciphertext modulus $q$.
Let $\mathcal{R}_q = \mathbb{Z}_q[X]/(X^N + 1)$ be the cyclotomic polynomial ring with a power-of-two polynomial ring degree $N$.
The client samples a secret polynomial $s(X) = s_0 + s_1X + \cdots +s_{N-1}X^{N-1}$, where each $s_i$ is sampled from a uniform ternary distribution in $\{-1, 0, 1\}$.
RLWE enables encrypting a plaintext polynomial $m(X) = m_0 + m_1X + \cdots + m_{N-1}X^{N-1}\in \mathcal{R}_p$ in a ciphertext.
Thus, we can put $N\log p$ bits of information in a single ciphertext.
For encrypting $m(X)$, the client samples a random polynomial $a(X)\in\mathcal{R}_q$, whose coefficients are sampled from a uniform distribution in $\mathbb{Z}_q$, and an error polynomial $e(X) \in \mathbb{Z}[X]/(X^N+1)$, whose coefficients are sampled from a discrete Gaussian distribution with a standard deviation $\sigma$.
Then, the client computes the following:
\[
\mathrm{Encrypt}(m(X)) \rightarrow \mathtt{ct}=(a(X),b(X)) \in \mathcal{R}_q^2
\]
\[
a(X)\cdot s(X) +b(X) = \Delta \cdot m(X) + e(X) \pmod{q}
\]
Here, $\Delta$ is determined as $\Delta= \lfloor \frac{q}{p} \rfloor$ (floored result).

We define the phase of an RLWE ciphertext in the similar way as LWE:
\[
\phi(\mathtt{ct})= a(X)\cdot s(X) +b(X) = \Delta \cdot m(X) + e(X)  \pmod{q},
\]
where each coefficient of the phase is in $\left[-\frac{q-1}{2}, \frac{q-1}{2}\right]$ range.

The client can recover the plaintext $m(X)$ with its coefficients in $\left[ -\frac{p-1}{2}, \frac{p-1}{2} \right]$ range by computing
\[
\mathrm{Decrypt}(\mathtt{ct})=\mathrm{round}\left(\frac{1}{\Delta}\cdot \phi(\mathtt{ct})\right),
\]
given that each coefficient of $e(X)$ is in $\left(-\frac{\Delta}{2}, \frac{\Delta}{2} \right)$ range.
 
Inter-ciphertext additions and plaintext-ciphertext multiplications are performed by polynomial additions and polynomial multiplications, respectively.
For $\mathtt{ct}=(a(X), b(X))$ and $\mathtt{ct}'=(a'(X), b'(X))$, they are computed as follows:
\[
\mathtt{ct} +  \mathtt{ct}' = (a(X) + a'(X), b(X) + b'(X))
\]
\[
c(X) \cdot \mathtt{ct} = (c(X) \cdot a(X), c(X) \cdot b(X))
\]

In this paper, we use $p=2^{18}$, $q\simeq2^{87}$ ($\text{\#RNS}=3$), $\sigma=3.2$, $N=2^{12}$, and $\Delta=\lfloor\frac{q}{p}\rfloor\simeq 2^{69}$ as the default parameters.
However, for our implementation of the OnionPIRv2~\cite{iacr-2025-onionpirv2} protocol, we use $p=2^{32}$, $q\simeq2^{108}$ ($\text{\#RNS}=4$), and $\Delta\simeq2^{76}$.

\subsection{Notations and symbols}

Table~\ref{tab:notation} summarizes the major notations and symbols used in the main text of this paper.

\setlength{\tabcolsep}{4pt}
\begin{table}[ht]
\caption{Notations and symbols.}
\label{tab:notation}
\begin{tabularx}{0.99\columnwidth}{lX}
\toprule
Symb. & Explanation\\
\midrule
\DB & Database.\\
$D_i$ & $i$-th \DB dimension.\\
$n$ & LWE dimension ($1{,}280$ throughout this paper).\\
$\mathpolyring$ & Polynomial ring $\mathbb{Z}_q[X]/(X^N+1)$.\\
$N$ & Degree of $\mathpolyring$ (typically $2^{12}$).\\
$q$ & Modulus of $\mathpolyring$.\\
$q_i$ & $i$-th RNS prime composing $q$.\\
\#RNS & The number of RNS primes composing $q$.\\
\bottomrule
\end{tabularx}
\end{table}
\setlength{\tabcolsep}{6pt}

\subsection{RNS prime selection}

When selecting the RNS primes ($q_i$'s), we make sure that
\[
q_i=1 \bmod 2N,
\]
which is a sufficient condition for enabling NTT.

Further, we make sure that $2N$ divides the RLWE plaintext modulus $p$ and
\[
q_i=1 \bmod p.
\]
This RNS prime selection minimizes the errors from RLWE because
\[
\Delta \cdot p + 1 = q\ \text{ for scale }\ \Delta=\lfloor\frac{q}{p}\rfloor.
\]
With this setting, overflows/downflows in the plaintext domain only result in small error increases.
For example, $\frac{3}{8}p + \frac{3}{8}p \rightarrow -\frac{1}{4}p$ in $\mathcal{R}_p$ translates to:
\[
\Delta \cdot \frac{3}{8}p + \Delta \cdot \frac{3}{8}p = \Delta \cdot \frac{3}{4}p = \Delta \cdot \left(-\frac{1}{4}p\right) - 1 \pmod{q}  
\]

For the following sections, we ignore such errors in RLWE ciphertexts that stem from overflows/downflows in the plaintext domain, as our RNS prime selection make them have minimal impact on precision; errors from other sources (e.g., encryption) are much greater.
Then, we can calculate the phase of inter-ciphertext addition and plaintext-ciphertext multiplication results of RLWE ciphertexts as follows:
\begin{align*}
\phi(\mathtt{ct} + \mathtt{ct}') &= \Delta \cdot (m(X) + m'(X)) +
(e(X) + e'(X))\\
&= \phi(\mathtt{ct}) + \phi(\mathtt{ct}')\\
\phi(c(X) \cdot \mathtt{ct}) &= \Delta \cdot (c(X) \cdot m(X)) +
(c(X) \cdot e(X))\\
&= c(X) \cdot \phi(\mathtt{ct})
\end{align*}

We can observe that the phase function $\phi$ has a linear property for both LWE and RLWE ciphertexts when ignoring the plaintext-domain overflows/downflows.

Also, we select primes that are smaller than $2^{29}$.
This leaves extra room for numbers beyond the typical ranges for numbers in $\mathbb{Z}_{q_i}$, specifically $\left[-\frac{q_i - 1}{2}, \frac{q_i - 1}{2} \right]$ or $\left[0, q_i - 1 \right]$, which allows applying modular reductions in a lazy manner~\cite{tches-2021-100x, asplos-2026-cheddar}.

\section{Experimental Materials Utilized}
\label{app:experimental}

For the evaluation of prior studies, we mainly utilized their open-source implementations listed in Table~\ref{tab:github_source}.
We used the main branch of each repository as of April 1st, 2026.

\setlength{\tabcolsep}{2pt}
\begin{table}[ht]
    \centering
    \caption{Prior PIR studies and their data sources.}
    \label{tab:github_source}
    {
    \small
    \begin{tabularx}{0.99\linewidth}{X|l}
        \toprule
        Prior studies & Data source (GitHub) \\
        \midrule
        Spiral~\cite{sp-2022-spiral} & menonsamir/spiral \\
        OnionPIRv2~\cite{ccs-2021-onionpir, iacr-2025-onionpirv2} & chenyue42/OnionPIRv2\\
        HintlessPIR~\cite{crypto-2024-hintless} & google/hintless\_pir\\
        Simple/DoublePIR~\cite{usenixsec-2023-simplepir} \& YPIR~\cite{usenixsec-2024-ypir} & menonsamir/ypir \\
        InsPIRe~\cite{iacr-2025-inspire} & google/private-membership \\
        PIRonGPU~\cite{github-PIRonGPU} & Alisah-Ozcan/PIRonGPU \\
        \bottomrule
    \end{tabularx}
    }
\end{table}
\setlength{\tabcolsep}{6pt}

All implementations were evaluated using their default library parameter configurations unless otherwise specified. We consider \DB containing $D$ records in total.
For OnionPIRv2, we used $D_0=2^9$. $N=2^{12}$, $q\simeq2^{108}$, and plaintext modulus $p=2^{32}$, which provides lower latency than the default configuration.
For HintlessPIR, SimplePIR, DoublePIR, and YPIR, we set $D_0\approx\sqrt{D}$ following their default configurations.
For InsPIRe, we selected $D_0$ which minimizes communication cost and also evaluated various $D_0$ values for Figure~\ref{fig:intro}: $D_0\in \{2^{10}, 2^{11}, 2^{12}\}$ for 1GiB \DB, $D_0\in \{2^{12}, 2^{13}, 2^{14}\}$ for 4GiB \DB, and $D_0\in \{2^{14},2^{15},2^{16}\}$ for 16GiB \DB.
Increasing $D_0$ reduces execution time but increases communication cost, reflecting the trade-off inherent in the protocol.
PIRonGPU supports various choices of $N$ with significantly different performance; we used $N=2^{15}$, which yields the lowest execution time.


\section{Formal Description of ExpPack}
\label{app:exppack_details}

We provide the formal specification of our expansion-based ring packing (ExpPack) method, detailing the conversion from scalar-HE (LWE) ciphertexts to poly-HE (RLWE) ciphertexts.

For simplicity, suppose we want to convert a collection of $N$ LWE ciphertexts to a single RLWE ciphertext, where each LWE ciphertext encrypts a scalar value $m_i$ ($0\le i < N$) under the same secret vector $\mathbf{s}=(s_0, s_1, \cdots, s_{n-1})\in\{-1,0,1\}^n$.
These LWE ciphertexts can be collectively represented as $\mathtt{ct}_\text{pack}=(\mathbf{A, b})$, where $\mathbf{A} \in \mathbb{Z}_{p}^{N \times n}$ and $\mathbf{b} \in \mathbb{Z}_{p}^N$.
We are using $p$ as the LWE ciphertext modulus and the RLWE plaintext modulus at the same time.

This $\mathtt{ct}_\text{pack}$ is constructed by setting the $\mathbf{a}$-part of the $i$-th LWE ciphertext as the $i$-th row of $\mathbf{a}$, and the $b$-part as the $i$-th element of $\mathbf{b}$.
Then, we can similarly define the phase of this collection of LWE ciphertexts as:
\begin{align*}
& \phi(\mathtt{ct}_\text{pack})=\mathbf{A}\cdot \mathbf{s} + \mathbf{b} =\\
& (\Delta\cdot m_0 + e_0, \Delta\cdot m_1 + e_1, \cdots, \Delta\cdot m_{N-1} + e_{N-1}),\quad |e_i|<\frac{\Delta}{2}.
\end{align*}

\subsection{Tiptoe-style ring packing}

To facilitate packing, we perform a column-wise decomposition of the LWE matrix $\mathbf{A}$ as $\mathbf{A} = [\mathbf{A}_0 \mid \mathbf{A}_1 \mid \cdots \mid \mathbf{A}_{n-1}]$. This allows us to rewrite the phase as a summation:
\begin{equation}
\label{eq:lwe-decryption-column}
\phi(\mathtt{ct}_\text{pack})=\mathbf{b} + \sum_{i=0}^{n-1} \mathbf{A}_i \cdot s_i = \mathbf{m} \pmod{p}
\end{equation}

Tiptoe~\cite{sosp-2023-tiptoe} leverages this linear structure for packing, utilizing a method previously established in the context of column-based ring packing~\cite{asia-2017-column-ring-pack}. Under this construction, the client prepares $n$ independent RLWE ciphertexts, each encrypting one element of the secret vector $\mathbf{s}$:
\[
\mathtt{ct}_{\text{Pack},i} \gets \mathrm{Encrypt}(s_i).
\]

We define an isomorphism $\iota$ that maps a length-$N$ vector over $\mathbb{Z}_{p}$ to a polynomial in the ring $\mathcal{R}_{p}$ by treating the vector elements as coefficients:
\[
\iota: \mathbb{Z}_{p}^N \ni (a_0, a_1, \dots, a_{N-1}) \mapsto \sum_{j=0}^{N-1} a_j X^j \in \mathcal{R}_{p}
\]

By selecting RLWE parameters such that the plaintext domain is $\mathcal{R}_{p}$ and the ciphertext domain is $\mathcal{R}_q$ (where $q$ is the RLWE ciphertext modulus), the server can perform a homomorphic evaluation of Eq.~\ref{eq:lwe-decryption-column}. The computation is formulated as:
\begin{equation}
\label{eq:tiptoe-rp}
(0, \Delta \cdot \iota(\mathbf{b})) + \sum_{i=0}^{n-1} \iota(\mathbf{A}_i) \cdot \mathtt{ct}_{\text{Pack},i} \rightarrow \mathtt{ct}_\mathrm{res}
\end{equation}

In this expression, $(0, \Delta \cdot \iota(\mathbf{b}))$ represents a trivial RLWE encryption used to incorporate the offset vector $\mathbf{b}$ into the result; it is trivial that $\phi((0, \Delta \cdot \iota(\mathbf{b}))) = \Delta \cdot \iota(\mathbf{b})$.
The terms $\iota(\mathbf{A}_i) \cdot \mathtt{ct}_{\text{Pack},i}$ denote plaintext-ciphertext multiplications.
Because the phase computation and $\iota$ has a linear property, the phase of $\mathtt{ct}_\mathrm{res}$ becomes:
\begin{align*}
\phi(\mathtt{ct}_\mathrm{res}) &= \Delta \cdot \iota(\mathbf{b}) + \sum_{i=0}^{n-1} \iota(\mathbf{A}_i) \cdot (\Delta \cdot s_i + e_i(X))\\
&= \Delta \cdot \left(\iota(\mathbf{b})+\sum_{i=0}^{n-1} \iota(\mathbf{A_i}) \cdot s_i\right) + \sum_{i=0}^{n - 1} \iota(\mathbf{A}_i) \cdot e_i(X)\\
&= \Delta \cdot\iota\left(\mathbf{b}+\sum_{i=0}^{n-1} \mathbf{A_i} \cdot s_i\right) + \sum_{i=0}^{n - 1} \iota(\mathbf{A}_i) \cdot e_i(X)\\
&= \Delta \cdot \iota(\mathbf{m}) + \sum_{i=0}^{n - 1} \iota(\mathbf{A}_i) \cdot e_i(X)
\end{align*}
Thus, $\mathtt{ct}_\mathrm{res}$ can be decrypted to obtain $\iota(\mathbf{m})$ when the sum of the error terms ($\iota(\mathbf{A}_i) \cdot e_i(X)$) is small enough.

To ensure computational efficiency, the server lifts the coefficients of $\iota(\mathbf{A}_i)$ and $\iota(\mathbf{b})$ from $\mathbb{Z}_{p}$ to $\mathbb{Z}_q$ and transforms them into the NTT domain.
The resulting process enables the server to pack multiple scalar LWE instances into a single RLWE ciphertext as illustrated in Figure~\ref{fig:simplepir}.

\subsection{Expansion-based packing (ExpPack)}

To minimize communication overhead, ExpPack replaces the requirement for $n$ individual ciphertexts with a single RLWE ciphertext.
The client constructs an RLWE ciphertext $\mathtt{ct}_\text{ExpPack}$ by encoding the secret scalars $s_i$ as coefficients of a single polynomial:
\begin{equation}
\mathtt{ct}_\text{ExpPack} \gets \mathrm{Encrypt}\left(\sum_{i=0}^{n-1} s_i X^i\right)
\end{equation}
Note that the specific assignment of $s_i$ to coefficients may vary depending on the chosen expansion schedule (e.g., bit-reversed ordering). 

To recover the individual encryptions, the server applies the recursive expansion process described in SealPIR~\cite{sp-2018-sealpir}. A known property of this expansion method is that it yields ciphertexts encrypting $2^{\lceil\log_2 n \rceil} \cdot s_i$ rather than $s_i$ directly. To normalize these values, the server pre-scales the initial ciphertext $\mathtt{ct}_\text{ExpPack}$ by the modular inverse $2^{-\lceil\log_2 n \rceil} \pmod{q}$ prior to the expansion. 
While $2^{-\lceil \log_2 n \rceil}$ is a large integer in $\mathbb{Z}_q$, it is canceled out by the $2^{\lceil \log_2 n \rceil}$ scaling during expansion.

\section{Security \& Precision}
\label{app:security}

The security of our protocols follows directly from the security of their underlying components; we refer to SimplePIR~\cite{usenixsec-2023-simplepir}, SealPIR~\cite{sp-2018-sealpir}, and Tiptoe~\cite{sosp-2023-tiptoe} for detailed analyses.

For precision, we require a more careful error analysis as HE errors may propagate across PIR components.

\subsection{Error isolation via reinterpretation and ExpPack}

When performing reinterpretation, encrypted data is treated as a stream of plaintexts that are directly recovered by the client via decryption.
As a result, prior context about errors is lost, and previously accumulated errors are not propagated to subsequent steps.
As a result, precision analysis can be conducted independently for each segment separated by reinterpretation, most of which have already been analyzed in SimplePIR~\cite{usenixsec-2023-simplepir}.

Also, our ExpPack construction similarly treats scalar-HE ciphertexts as plaintexts (see Eq.~\ref{eq:tiptoe-rp}), making packing errors largely independent of previously accumulated errors.

Thus, the only remaining path that requires verification is ExpPack followed by the poly-HE dimension folding steps in our hybrid protocol.

\subsection{Error analysis of scalar-HE rounding}

Before performing the main computations of ExpPack, we reduce the ciphertext modulus of scalar-HE ciphertexts from $2^{32}$ to $p=2^{18}$, following Tiptoe~\cite{sosp-2023-tiptoe}.
This is equivalent to introducing an additional rounding error term into each element of a scalar-HE ciphertext $(a_0, a_1, \cdots, a_{n-1}, b)\in\mathbb{Z}_{2^{32}}^{n+1}$, which originally decrypts as
\[
b \cdot s_n + \sum_{i=0}^{n-1} a_i \cdot s_i = \Delta \cdot m + e.
\]
We are using a ternary secret $(s_0,\dots,s_{n-1})$, and a scale $\Delta = 2^{24}$.
We assume the previously accumulated error is sufficiently small: $|e| < \frac{\Delta}{4}$.

Let $r_{\mathbf{a},i}$ (for $0 \le i < n$) and $r_b$ denote the rounding errors for $a_i$ and $b$, respectively, where $|r_i| < 2^{13}$.
Each rounding error is in range $(-\frac{2^{31}}{p}, \frac{2^{31}}{p})$

The total additional error becomes
\[
E_\text{round} = r_b + \sum_{i=0}^{n} r_{\mathbf{a},i} \cdot s_i.
\]
To ensure correctness, it suffices to guarantee
\[
\left|E_\text{round}\right| < \frac{\Delta}{4}.
\]

Assuming that these rounding errors are i.i.d. uniform over $(-\frac{2^{31}}{p}, \frac{2^{31}}{p})$, where $s_i$ values are also i.i.d. and is $-1$, $0$, or $1$, we can bound the tail probability using Hoeffding’s inequality.
Each term lies in an interval of width $\frac{2^{32}}{p}$, and there are $(n+1)$ terms. Thus,
\[
\Pr\left[\left| E_\text{round} \right| \ge t \right]
\le 2 \exp\left( -\frac{2 t^2}{(n+1)\cdot (2^{32}/{p})^2} \right).
\]

For this LWE rounding, we use parameters $p=2^{18}$, which is also used as the RLWE plaintext modulus, $n=1{,}280$, and $\Delta=2^{24}$ (LWE scale).
When we plug in these values with $t = \frac{1}{4}\Delta = 2^{22}$, we obtain
\[
\Pr\left[\left| E_\text{round}\right| \ge \frac{\Delta}{4} \right]
\le 2 \exp(-102) < 2^{-146},
\]
which is negligible.
Even if we have trillion ($2^{40}$) ciphertexts, the union bound probability ($<2^{-106}$) is still negligible.

\subsection{Error analysis of ExpPack}

For ExpPack, we show that the maximum (not probabilistic) error remains sufficiently small.
As mentioned before, the error from the input scalar-HE ciphertexts have minimal impact as we interpret them as plaintexts during packing.
Suppose the error of $\mathtt{ct}_{\text{Pack,i}}$ is bounded by $|E_\text{expand}|$.
As we accumulate $n$ plaintext-ciphertext multiplication results as shown in Eq.~\ref{eq:tiptoe-rp}, we can bound the total error to $n\cdot |E_\text{expand}|$.
We make this sufficiently small by setting the ciphertext modulus $q$ large enough (e.g., $q\simeq2^{87}$).

$|E_\text{expand}|$ is determined by the expansion process.
If the error of a freshly encrypted ciphertext is bounded by $6\sigma$ (typically $\sigma=3.2$), the maximum size of error from expansion is determined as follows~\cite{sp-2022-spiral}:
\[
|E_\text{expand}|< 2^{\lceil \log_2 n \rceil} \cdot (\ell t + 1) \cdot 6 \sigma.
\]

Here, $\ell$ is the decomposition degree for the expansion keys, and $t$ is determined as
\[
t = 2^{\lceil\frac{\log_2 q}{\ell}\rceil}.
\]
The parameters we use are $n=1{,}280$, $q\simeq2^{87}$, $\ell=5$, $t=2^{18}$, and $\sigma=3.2$.
Plugging in these values, we roughly get $|E_\text{expand}|<2^{35.6}$.
Thus, the maximum error from ExpPack with expansion is strictly bounded by:
\[
|E_\text{ExpPack}| < n|E_\text{expand}| < 2^{46},
\]
which does not affect the precision as it is much smaller than the poly-HE ciphertext scale $\Delta=\frac{q}{p}\simeq2^{69}$ for $p=2^{18}$.

\subsection{Error in poly-HE folding for our hybrid protocol}

Although poly-HE folding steps directly follow ExpPack, the errors introduced during poly-HE folding are additive and do not depend on the errors already accumulated in the RLWE ciphertexts.
As we have shown that these prior errors remain sufficiently small relative to the scale $\Delta$, the remaining error analysis for poly-HE folding follows directly from OnionPIR~\cite{ccs-2021-onionpir}, which we defer a detailed analysis to.
The only difference is that the expansion process is shared with that used for the special ExpPack ciphertexts.

\balance

\subsection{Error analysis of RLWE modulus switching}

At the end of each packing (except in our hybrid protocol, where it is deferred to the end of the online server computation), we perform modulus switching on the resulting RLWE ciphertexts, reducing the modulus from $q$ to approximately $2^{29}$, corresponding to ${\text{\#RNS}=1}$.

This modulus reduction has similar effects to scalar-HE rounding.
For an RLWE ciphertext $\mathtt{ct}=(a(X) = a_0 + a_1X +\cdots + a_{N-1}\cdot X^{N-1}, b(X) = b_0 + b_1X +\cdots + b_{N-1}\cdot X^{N-1})$, which decrypts as:
\[
a(X) \cdot s(X) + b(X) = (\Delta\cdot m_0 + e_0) + \cdots + (\Delta\cdot m_{N-1} + e_{N-1})X^{N-1} 
\]
for a ternary secret $s(X)=s_0+s_1X+\cdots +s_{N-1}X^{N-1}$ and scale $\Delta$, which is set to $\lfloor\frac{q}{p}\rfloor$ when coupled with scalar-HE rounding.

If we only look at the coefficient of $X^k$, the following holds:
\[
b_k + \sum_{i=0}^{N-1} a_i \cdot s_{(k-i)\bmod N} = \Delta \cdot m_k + e_k.
\]
Assume that the previously accumulated error is sufficiently small: $|e_k|<\frac{\Delta}{4}$.

When reducing the modulus from $q$ to $2^{29}$, each coefficient of $a(X)$ and $b(X)$ incurs a rounding error. Let $r_{a,i}$ denote the rounding error for $a_i$ and $r_{b,i}$ for $b_i$, where $|r_{a, i}|, |r_{b,i}| < \frac{q}{2^{30}}$.
The resulting additional error for the $k$-th coefficient becomes
\[
E_\text{round} = r_{b, i} + \sum_{i=0}^{N-1} r_{a, i} \cdot s_{(k-i)\bmod N}.
\]

To guarantee correct decryption, it suffices to ensure
\[
\left| E_\text{round} \right| < \frac{\Delta}{4}.
\]

Assuming the rounding errors are i.i.d.\ and uniformly distributed in $\left(-\frac{q}{2^{30}}, \frac{q}{2^{30}}\right)$, we can bound the tail probability using Hoeffding’s inequality. Since each term is bounded within an interval of width $\frac{q}{2^{29}}$ and there are $(N+1)$ terms, we have
\[
\Pr\left[\left| E_\text{round} \right| \ge t \right]
\le 2 \exp\left(-\frac{2t^2}{(N+1)\cdot (q/2^{29})^2}\right).
\]

By plugging in our parameters, which are $N=2^{12}$, $p=2^{18}$, $\Delta=\lfloor \frac{q}{p}\rfloor$, and $t=\frac{\Delta}{4}$, we get:
\[
\Pr\left[\left| E_\text{round} \right| \ge \frac{\Delta}{4} \right] \le 2\exp(-128) < 2^{-183}.
\]
Even if we consider all the $N=2^{12}$ coefficients and trillion ($2^{40}$) ciphertexts, the union bound probability ($<2^{-131}$) remains negligible.

\end{document}